\documentclass[pdflatex,sn-aps]{sn-jnl}

\usepackage{graphicx}%
\usepackage{multirow}%
\usepackage{amsmath,amssymb,amsfonts}%
\usepackage{amsthm}%
\usepackage{mathrsfs}%
\usepackage[title]{appendix}%
\usepackage{color}
\usepackage{xcolor}%
\usepackage{textcomp}%
\usepackage{manyfoot}%
\usepackage{booktabs}%
\usepackage{algorithm}%
\usepackage{algorithmicx}%
\usepackage{algpseudocode}%
\usepackage{listings}%
\usepackage{comment}

\usepackage{eqnarray,amsmath}
\usepackage{dcolumn}
\usepackage{ragged2e}
\usepackage{array}
\usepackage{etoolbox}
\usepackage{multirow}
\usepackage{anyfontsize}

\usepackage[]{bibunits}
\usepackage{hyperref}

\usepackage[sort&compress]{natbib} 
\setcitestyle{numbers}
\makeatletter
\newcommand*{\newbibstartnumber}[1]{%
  \apptocmd{\thebibliography}{%
    \global\c@NAT@ctr #1\relax
    \addtocounter{NAT@ctr}{-1}%
  }{}{}%
}
\makeatother

\counterwithin{figure}{section}
\renewcommand{\thefigure}{\arabic{figure}}

\counterwithin{table}{section}
\renewcommand{\thetable}{\arabic{table}}

\begin{document}
\begin{bibunit}[sn-nature] 


\title[Article Title]{
\LARGE{\textbf{An Alternate Pathway for H$_2$ Formation}} 

\LARGE{\textbf{in the Early Universe:}}

\vspace{5pt}
\large{\textbf{A physical process to account for the presence and coevolution of the luminous galaxies and supermassive black holes at the high redshifts}}
}

\author*[]{\fnm{Amrendra} \sur{Pandey}}\email{amrendra.pandey@universite-paris-saclay.fr}
\author[]{\fnm{Olivier} \sur{Dulieu}}\email{olivier.dulieu@cnrs.fr}
\author[]{\fnm{Nadia} \sur{Bouloufa-Maafa}}\email{nadia.bouloufa@universite-paris-saclay.fr}

\affil[]{\orgdiv{Laboratoire Aim$\acute{\text e}$ Cotton}, \orgname{Universit\'e Paris-Saclay, CNRS}, \orgaddress{\city{Orsay}, \postcode{91405}, \state{Orsay}, \country{France}}}

\abstract{\textbf{Molecular hydrogen (H$_2$) and hydrogen deuteride (HD) are key coolants in primordial gas and regulate the formation of the first stars and proto-galaxies. Recent results from the James Webb Space Telescope provide striking insights into galaxies detected at high redshifts, which are found to be significantly more abundant and luminous than expected from galaxy formation models, thus suggesting a gap in our understanding of the early Universe. Standard pathways for H$_2$ formation in the early Universe proceed through the H$^-$ and H$_2^+$ intermediates, both of which are strongly suppressed at high redshift by the cosmic microwave background. We propose an additional pathway for H$_2$ and HD formation that could be active as early as the end of the epoch of recombination and could enable the formation of the first stars earlier than the current prediction at redshift $z \simeq 30 - 20$. The proposed pathway relies on the manifestation of Jahn-Teller dynamical coupling between electronic states of H$_3^+$. This coupling induces transient three-body recombination in H$^+$, H and H, and charge exchange within the charged atom-dimer complex (H$+$H$_2^+$ or H$^+$$+$H$_2$) that directly creates ground-state H$_2$ (and HD), bypassing the fragile intermediates that limit the standard primordial pathways. Our analysis shows that this mechanism could occur under the thermodynamic conditions of the post-recombination epoch, also suggesting that it might be playing a role in the active galactic nuclei feedback processes, regulating the formation rates of the first stars and the accretion rates of the first black holes. Though the global impact on galaxy formation and black-hole growth is not yet determined and will require quantitative assessment in future modeling, the mechanism offers an additional chemical route for H$_2$ and HD formation, with substantial cosmological relevance for primordial chemistry and early structure formation.}}

\maketitle

\section{Introduction}\label{s1int} 

Understanding the formation of the first stars is a central issue in modern astrophysics and cosmology \cite{barkana2001beginning, bromm2009formation, bromm2013formation}. In the standard cold dark matter model, the first generation of stars, known as Population III (Pop III), formed from primordial, metal-free gas in the early Universe at redshifts $z \simeq 30-20$. These stars emerged inside dark matter minihalos with typical masses of $\sim 10^{6}$ M$_\odot$ (M$_\odot$ is the solar mass) and virial temperatures of order 1,000 K \cite{bromm2009formation, tegmark1997small, bromm2013formation, klessen2023first, greif2015numerical}.
Star formation in such environments requires the collapsing gas to cool efficiently. This condition is captured by the Rees-Ostriker-Silk criterion, which specifies when the radiative cooling in a dark matter minihalo is sufficient to allow gas fragmentation and the onset of star formation, marking the beginning of galactic evolution \cite{bromm2013formation}.
From the end of the recombination era ($z \sim 1100$), through the Dark Ages, and into the epoch of reionization (down to $z \sim 6$), cooling in the absence of heavy elements depended almost entirely on molecular hydrogen (H$_2$) \cite{saslaw1967molecular, peebles1968origin}. Consequently, the chemical evolution of H$_2$ played a decisive role in controlling the thermal state of the gas and the formation of the first stars \cite{galli2013dawn, klessen2023first}.

Current cosmological models indicate that the Big Bang nucleosynthesis created mostly $^1$H ($\sim 75$\%) and $^4$He ($\sim 25$\%), as well as about 0.01\% of $^2$D and $^3$He, and a trace fraction consisting of other light elements, such as lithium and beryllium.
Before the epoch of reionization, cooling at $\sim$ 10$^4$ K was mediated by atomic H. 
At lower temperatures, the gas cooled through H$_2$ with cooling rate that scales roughly linearly with its fraction and the excitation energy of its lowest rovibrational transition, reaching a minimum temperature $\sim$ 200 K at a density of $n_\text{H} \sim$ $10^4$ cm$^{-3}$ (Pop III.1 formation pathway) \cite{klessen2023first}. Numerical simulations based on the primordial H$_2$ chemistry and the associated cooling and heating processes suggest that the first stars were massive ($\gtrsim$ 100 M$_\odot$) and short-lived ($\sim$ few Myr) \cite{klessen2023first, bromm2013formation, bromm2009formation, tegmark1997small}. 
In conditions where the HD population increases, the primordial gas can cool even further via HD, reaching temperatures of approximately 50-100 K at a density of $n_\text{H} \sim$ $10^6$ cm$^{-3}$. This enhanced cooling enables the formation of relatively low-mass stars ($\gtrsim$1-10 M$_\odot$ with lifetimes of $\sim 10$s Myr) in dark matter mini-halos that are smaller by about an order of magnitude compared to those associated with H$_2$-dominated cooling (Pop III.2 formation pathway) \cite{klessen2023first}.
The intrinsic distinction between these two scenarios is that HD can cool the gas via the dipole-allowed rotational transition $J_{\text{HD}} = 1 \rightarrow 0$ with lower excitation energy of $\simeq 128$ K, whereas H$_2$ cooling relies on the much slower quadrupole rotational transition $J_{\text{H}_2} = 2 \rightarrow 0$ with an excitation energy of $\simeq 512$ K.

James Webb Space Telescope (JWST) observations have revealed very luminous galaxies at Cosmic Dawn ($z> 10$), pointing to extremely rapid galaxy formation \cite{bunker2023jades, robertson2024earliest, carniani2024spectroscopic, zavala2025luminous, naidu2025cosmic, adamo2025first}. Spectroscopically confirmed objects such as JADES-GS-z14-0 ($z \approx 14$) \cite{carniani2024spectroscopic} and MoM-z14 ($z = 14.44$) \cite{naidu2025cosmic} show that luminous galaxies already existed less than 300 million years after the Big Bang, with an abundance well above pre-JWST expectations.
Several other sources (e.g., GN-z11 \cite{bunker2023jades}, GHZ2/GLASS-z12 \cite{zavala2025luminous}) exhibit unusual chemical abundance patterns, including elevated N/O ratios, suggesting enrichment dominated by very massive stars. The detection of [O III] 88 $\mu$m emission in JADES-GS-z14-0 \cite{carniani2025eventful} further indicates early oxygen enrichment of the interstellar medium.
Other studies, \cite{xiao2024accelerated}, report massive, dust-obscured galaxies at $z =  5-9$ with baryon-to-stellar conversion efficiencies approaching $\sim 50$\%, far higher than those inferred for the most efficient systems at later epochs. In addition, candidate galaxies at $z \approx 17-25$ \cite{perez2025rise}, if spectroscopically confirmed, would imply even more extreme early growth, which will be difficult to reconcile with standard galaxy-formation models. Potential explanations include enhanced star-formation efficiency or strong dust attenuation, a top-heavy initial mass function, or contributions from active galactic nuclei \cite{mason2023brightest, ferrara2023stunning, trinca2024exploring}.
Furthermore, observations of spiral galaxies such as ceers-2112 \cite{costantin2023milky}, Big Wheel \cite{wang2025giant}, and Alaknanda \cite{jain2025grand} from the first two billion years of the Universe challenge traditional models of galaxy evolution. In short, along with the observation of luminous galaxies at $z > 10$ and their earlier chemical enrichment, the presence of supermassive black holes in the first few billion years of the Universe \cite{fan2023quasars, inayoshi2020assembly, volonteri2021origins, smith2019supermassive, jeon2025physical}, detection of the large scale structures like spiral and bar formation in the early galaxies suggests towards the timeline shift of the star formation, more efficient stellar and cosmic structure evolution.

In the era before the Cosmic Dawn ($z > 10$), molecular hydrogen existed only in trace amounts. Its relative abundance, $x\text{H}_2$ ($= n_{\text{H}_2}/n_\text{H}$), was approximately $10^{-6}$ until $z \sim 100$, and eventually reaching $10^{-3}$ by $z \simeq 30 - 20$, primarily due to H$^-$ \cite{peebles1968origin} and H$_2^+$ \cite{saslaw1967molecular} mechanisms in collapsing dark matter minihalos \cite{klessen2023first, bromm2013formation}. Both H$^-$ and H$_2^+$ mechanisms are two-step reactions and are limited at high redshifts. In cosmological simulations, the H$^-$ mechanism dominates at redshifts of $z \simeq 30 - 20$, when the first stars formed \cite{hirata2006cosmological}. A third process, i.e., three-body reaction among three neutral hydrogen atoms, becomes important at high particle densities above $\sim$ 10$^9$ cm$^{-3}$ \cite{galli2013dawn, klessen2023first}. These reactions and their coupling with other trace atomic species in the early Universe have been widely studied  \cite{galli2013dawn, hirata2006cosmological}. Today ($z = 0$), H$_2$ mainly forms by the grain surface catalytic process (H $+$ H $+$ surface $\rightarrow$ H$_2$ $+$ surface) \cite{grieco2023enhanced}.

In this work, we propose a new reaction pathway that produces neutral H$_2$ and HD in their ground electronic states. Based on a previous study of the ion-atom-atom rubidium system \cite{pandey2024ultracold}, the formation results from transient three-body recombination (TTBR) and charge exchange (CE) occurring on Jahn-Teller-induced (JT-induced) conical intersections \cite{yarkony1996diabolical} in a pair of singlet H$_3^+$ states of colliding H$^+$, H$(1s)$, and H$(1s)$ triads. This proposed mechanism likely influences the formation rates of H$_2$ and HD in the early Universe, with consequences for the onset and efficiency of structure formation and subsequent evolution.

\begin{figure}[t!]
\centering
\includegraphics[scale=0.345]{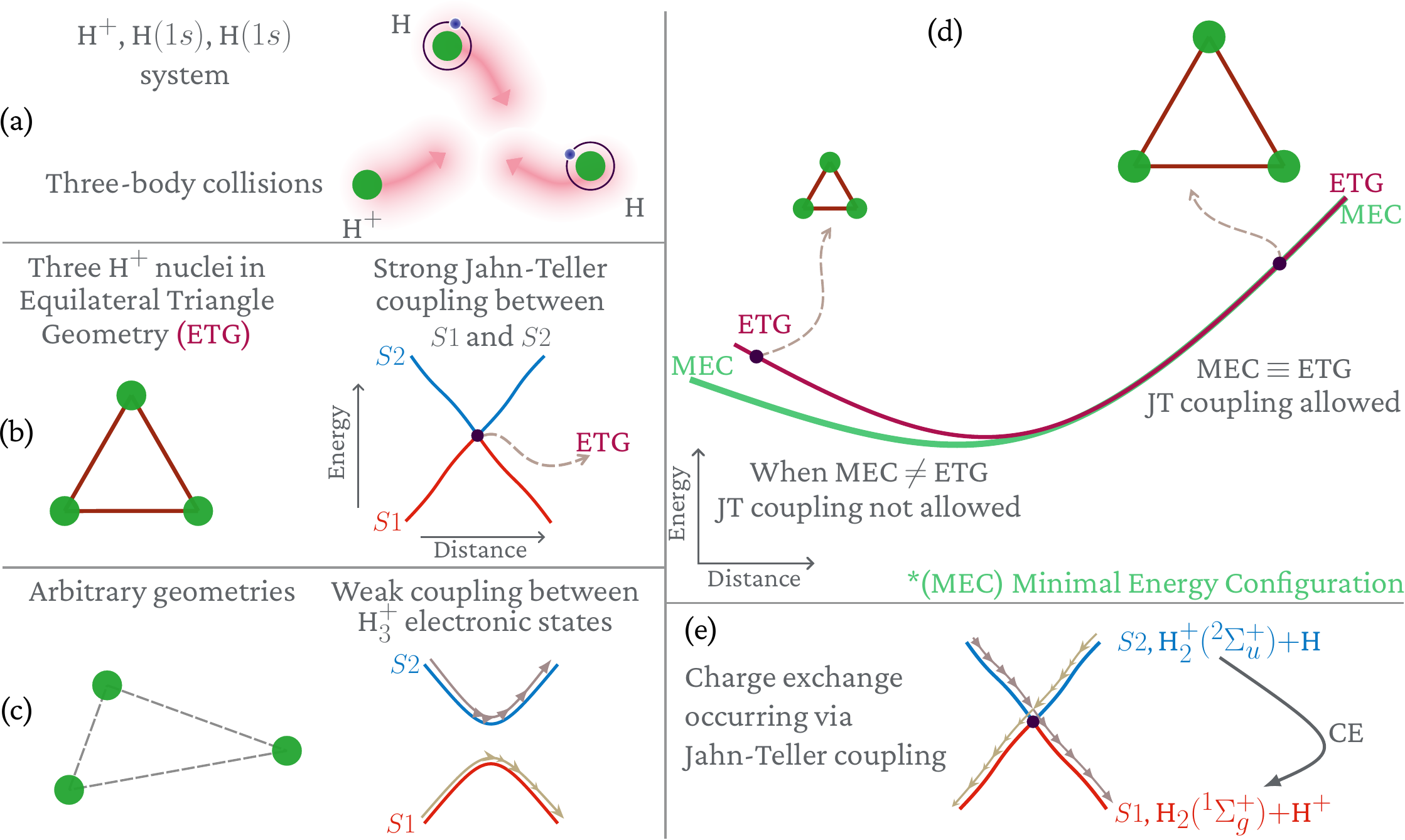}
\caption{$\vert$ \textbf{Jahn-Teller coupling in H$_3^+$ states.} (a) Schematic diagram of two hydrogen atoms, H$(1s)$, and a H$^+$ ion, interacting in a three-body collision via Coulomb forces. (b) Equilateral triangle geometry (ETG) for three H$^+$ nuclei along with the cuts to the two singlet potential energy surfaces (PESs) of the two-electron H$_3^+$ system, $S1$ and $S2$, depicted as functions of the internuclear separation, varying from left to right. 
In a collision, when the nuclei reach an ETG, the two singlet surfaces $S1$ and $S2$ cross each other through Jahn-Teller coupling. (c) At the geometries other than ETGs, $S1$ and $S2$ usually don't intersect, preventing curve-crossing reactions. 
The grey arrows on top of $S_1$ and $S_2$ represent movements of the nuclei guided by the PESs. (d) Two cuts to a PES of H$_3^+$. The curve labeled ETG represents the seam of the JT-coupled singlet states where $S1$ and $S2$ are degenerate. 
The other curve depicts the potential energies along the minimal energy configuration (MEC) and is representative of the geometries accessed by the colliding H$^+$, H$(1s)$, and H$(1s)$. JT coupling will be accessed in collisions only in the region where MEC is identical to the ETG, satisfying the primary constraint for the JT-induced reaction, i.e., TTBR and CE. (e) The possibility of CE is further controlled by the requirement that $S1$ and $S2$ be reciprocally associated with H$_2^+$ and H$_2$ dimers.}
\label{fig:JTsch}
\end{figure}

\section{A new reaction pathway for H$_2$ and HD formation} \label{s2res}

For homonuclear triatomic molecules like H$_3^+$, the collective nuclear Coulomb field of the high-symmetry equilateral triangle geometry (ETG), belonging to the $D_{3h}$ point group, leads to intersections in its electronic states, often termed symmetry-required conical intersections.
Jahn-Teller-induced conical intersections are ubiquitous to these triatomic molecules and have been studied for many alkali neutral and ionic species \cite{pandey2024ultracold}. 
For H$_3^+$, out of three singlet and three triplet electronic states associated with the first fully atomized asymptotic limit, H$^+$$+$H($1s$)$+$H($1s$), a pair of singlet states, $2^1A'$ and $3^1A'$, and a pair of triplet states, $1^3A'$ and $2^3A'$, exhibit Jahn-Teller conical intersections at the ETGs for all interparticle separations. JT coupling in the singlet states of H$_3^+$, namely $2^1A'$ (as $S1$) and $3^1A'$ (as $S2$), is shown using a few schematic diagrams in Fig. \ref{fig:JTsch} (a) - (c).
H$_2$ formation via the JT-induced TTBR+CE mechanism in $S1$ and $S2$ states involves JT-induced transient three-body recombination (TTBR) among ion-atom-atom H$^+$$+$H($1s$)$+$H($1s$), and JT-induced charge exchange (CE) of the transition state \{H$_2^+$($^2\Sigma_u^+$) $+$ H($1s$)\} to neutral H$_2$($X^1\Sigma_g^+$) and H$^+$, 

\begin{equation}
\text{H}^+ + \text{H}(1s) + \text{H}(1s) \xrightarrow{\text{TTBR}} \{\text{H}_2^+(^2\Sigma_u^+) + \text{H}(1s) \} \xrightarrow{\text{CE}} \text{H}_2(X^1\Sigma_g^+, v_g) + \text{H}^+.
\label{eq:eqH3ptbr}
\end{equation}

The requirement for the JT-induced TTBR is illustrated in Fig. \ref{fig:JTsch} (d): i.e., JT coupling between $S1$ and $S2$ will be accessed in collisions only in the region where the electronic potential energy for the minimal energy configuration (MEC) is identical to the one for ETGs. JT-induced CE requires one more condition to be fulfilled: i.e., the curves $S1$ and $S2$ must be reciprocally associated with H$_2^+$ and H$_2$ dimers (as illustrated in Fig. \ref{fig:JTsch} (e)). These arguments are developed in Methods (see details in section \ref{ssM1jt}) and include the necessary discussion of the computation of H$_3^+$ potential energy surfaces (PESs). The role of the participants' initial kinetic energies is also discussed in Methods (see details in section \ref{ssM3thrm}). In addition to H$_2$, the JT-induced TTBR+CE mechanism would also produce HD when the reaction occurs among H$^+$$+$H($1s$)$+$D($1s$) complex, given that under the Born-Oppenheimer approximation the PESs of H$_3^+$ and H$_2$D$^+$ are identical.

Since both JT-induced TTBR and CE reactions involve $S1$ and $S2$ H$_3^+$ states, they can occur in a single collision. This is a crucial difference from the two standard pathways for primordial H$_2$ formation (via H$^-$ and H$_2^+$ intermediates), as they are two-step processes. 
Their initial steps, namely the radiative attachment forming H$^-$ and the radiative association forming H$_2^+$, are highly susceptible to destruction through photodetachment of H$^-$ or photodissociation of H$_2^+$ by the intense thermal cosmic microwave background (CMB) radiation and by non-thermal recombination photons in the Wien tail \cite{hirata2006cosmological,  coppola2013non}. 
For this reason, both are limited at high redshift. H$_2$ production peaks at $z \sim 260$ for the H$_2^+$ mechanism and at $z \sim 90$ for the H$^-$ mechanism \cite{hirata2006cosmological}. 
In contrast, once H$_2$ in the ground electronic state $X^1\Sigma_g^+(v_g)$ is produced, it is relatively robust---the CMB and non-thermal photons are too weak at post-recombination redshifts to excite H$_2$ to the Lyman-Werner (LW) bands, making its photodissociation negligible \cite{hirata2006cosmological}. Uniquely, as the proposed JT-induced TTBR+CE mechanism forms ground-state H$_2$ directly in a single step, it bypasses the fragile H$^-$ and H$_2^+$ intermediates (further details in section \ref{ssD3supp}). 
Because this pathway relies on non-adiabatic couplings and rapid transitions near conical intersections, known to be efficient in molecular formation \cite{ferretti1997quantum, farfan2020systematic}, it is likely to provide an additional H$_2$ formation route under primordial conditions where the conventional channels are strongly suppressed. 
For instance, in the high H-density conditions with considerable H$^+$, i.e., at the end of the epoch of recombination ($z \sim 1100$), in the outflows of the active galactic nuclei (AGNs) hosting black holes, and in star-forming collapsing dark matter minihalos, the H$_2$ fraction likely rises significantly from the JT-induced TTBR+CE reactions. 
While the overall cosmological impact remains to be quantitatively assessed, this mechanism is expected to be active in the post-recombination epoch. As a promising trend, the motivation for the present study is rooted in our earlier investigations of heavier ion-atom-atom systems Rb$^+$$+$Rb$+$Rb \cite{pandey2024ultracold}, where explicit inclusion of Jahn-Teller coupling was shown to resolve apparently contradictory experimental observations of three-body recombination reactions. In those systems, the JT-induced mechanism provided a unified explanation for the simultaneous production of weakly bound ionic dimers and deeply bound neutral dimers, while reproducing experimentally observed reaction rates that could not be accounted for within conventional semi-classical three-body recombination models. These results demonstrate that Jahn-Teller coupling can qualitatively drive three-body collision dynamics and open efficient reaction pathways that would otherwise be inaccessible. Building on this insight, the present work extends the same conceptual framework to the prototypical and astrophysically most relevant system, i.e., H$^+$$+$H$(1s)$$+$H$(1s)$.

Additionally, Jahn-Teller couplings in the excited manifolds of H$_3^{+*}$ imply that reaction channels analogous to the TTBR+CE process in ground-state H$^+$$+$H($1s$)$+$H($1s$) can also operate when one or more H atoms are in excited states. 
These pathways can produce excited H$_2^\ast$, which subsequently relaxes to the ground electronic state through radiative or non-radiative processes, thereby contributing to H$_2$ formation in environments where a significant fraction of hydrogen resides in excited levels.
Such conditions may arise in astrophysical environments characterized by strong radiation fields or dense photoionized gas. For instance, some AGNs show Balmer line absorption features (e.g., H$_\alpha$, H$_\beta$), indicating the presence of H($n = 2$) in their line-of-sight gas \cite{juodvzbalis2024rosetta, maiolino2024diverse, matthee2024little, taylor2025capers, harrison2024observational, alexander2025drives}, and thus demonstrating the existence of pockets of excited atomic hydrogen under the high-energy conditions near accreting supermassive black holes.
Our results suggest that, if efficient, JT-induced TTBR+CE pathways for excited H$_3^{+*}$ manifolds could constitute an additional route for H$_2$ formation in localized regions of AGN outflows or irradiated gas. 
The enhanced cooling may, therefore, accelerate galactic growth further once the proto-galaxy has developed a central black hole. Observational evidence shows that AGN-driven winds can contain or generate molecular material \cite{riffel2020active, gallagher2019widespread}, and some studies suggest that molecules may form in situ within these outflows \cite{richings2018origin}. 
Considering these astrophysical environments, our alternate pathway promises to have a significant impact on the star formation processes in the early Universe.

\section{Early structure formation, the first stars and black holes} \label{ssD1impli}

The JT-induced TTBR+CE mechanism implies an enhancement in H$_2$ and HD fractions at early times, thereby modifying the cooling properties of gas within collapsing halos and, consequently, the characteristic mass scale of the first stars and black holes.
Specifically, in the extreme scenario where there is efficient production of HD at the end of the epoch of recombination ($z \sim 1100$), this would suggest a rapid stellar development at a very early stage. 
The smaller characteristic excitation energy of HD ($\sim$ 128 K) compared to H$_2$ ($\sim$ 512 K), and the resulting lower virial temperature (T$_{vir}$) of cooling gas, imply a lowering of the mass of dark-matter minihalos (M $\propto$ $(\text{T}_{vir}/(z + 1))^{3/2}$) \cite{barkana2001beginning, ripamonti2007role}) by about an order of magnitude compared to the H$_2$-dominated cooling-based Pop III.1 star formation pathway; for instance, from M $\sim 10^6$ M$_\odot$ to $\sim 10^5$ M$_\odot$. 
Consequently, for the HD-dominated cooled primordial gas, the smaller dark-matter minihalos will be assembled at the earlier time at redshift $z \sim 50$, instead of the typical case of Pop III.1 formation pathway at $z \approx 30 - 20$ \cite{mebane2018persistence}. 
As a result, it could provide an additional $\sim 100$ Myr for galactic evolution, substantially increasing current theoretical estimates that rely on the Pop III.1 formation pathway for the first stars.

Considering the synopsis of early formation of the first stars, efficient cooling from additional H$_2$ and HD would also influence the creation and growth of the first black holes (BHs) in proto-galaxies. 
Over the past decades, luminous quasars, powered by accretion onto supermassive black holes (SMBHs) with masses around $10^9 \text{M}_\odot$ at high redshifts ($z > 6$), have been observed. Additionally, JWST has detected fainter SMBHs at even higher redshifts, with masses in the range $10^6-10^8 \text{M}_\odot$. The presence of these objects in the early Universe has been challenging to understand \cite{fan2023quasars, inayoshi2020assembly, volonteri2021origins,  smith2019supermassive, jeon2025physical}. Many scenarios for the formation of the first BHs and the growth mechanisms for SMBHs have been proposed. 
Black hole seed models include light seeds from $\sim 100\text{M}_\odot$ Pop III.1 remnants at $z \gtrsim 20$ \cite{madau2001massive}, intermediate-mass seeds formed via stellar collisions or black hole mergers in dense clusters \cite{inayoshi2020assembly}, and heavy seeds produced as $\sim 10^5-10^6\text{M}_\odot$ direct-collapse black holes (DCBHs) from metal-free gas cooled only through atomic H in $\sim 10^8\text{M}_\odot$ dark matter halos with virial temperatures above $10^4$ K \cite{inayoshi2020assembly}. Despite extensive study, no consensus exists \cite{jeon2025physical}. Light-seed scenarios rely on sustained super-Eddington accretion \cite{lupi2024sustained}, which is disfavored by AGN feedback and star formation activities \cite{johnson2007aftermath}, while intermediate-mass and direct-collapse seeds appear too rare in simulations to explain the observed abundance of SMBHs \cite{o2025predicting}.

Efficient cooling and rapid fragmentation---caused by increased H$_2$ and HD production in the early Universe---undermine heavy-seed DCBH models. These models require suppression of molecular cooling to prevent gas fragmentation \cite{inayoshi2020assembly, smith2019supermassive}. Therefore, they consider the presence of intense LW irradiation, which would destroy the built-up H$_2$ population \cite{inayoshi2020assembly, smith2019supermassive}. On the contrary, if LW photons are present, they would also excite H($n = 2, 3 ...$) and promote H$_2$ formation through H$_3^{+*}$ excited states,  thereby further reducing the likelihood of direct collapse scenarios.

In contrast, efficient molecular cooling will raise interstellar gas density and enhance accretion, potentially reaching super-Eddington rates \cite{inayoshi2020assembly, smith2019supermassive}. 
Therefore, the increased H$_2$ and HD generated by the JT-induced TTBR+CE mechanism favor the formation of light-seed BHs and support their sustained growth by facilitating the accretion of dense, efficiently cooled stellar matter.
H$_2$ and HD production from excited H$_3^{+*}$ states may also help drive accretion in the first high-redshift AGNs \cite{juodvzbalis2024rosetta, maiolino2024diverse, matthee2024little, taylor2025capers, harrison2024observational, alexander2025drives}. Moreover, the timeline shift of the first stars, and consequently, the first black holes, possibly to redshift range $30 < z < 50$ (from the upper limit of Pop III.1 formation case to the proposed scenario of JT-induced TTBR+CE HD-dominated cooling), provides additional time up to $\sim 100$ Myr for growth, reducing the strain on the required accretion rates.
Observations of super-Eddington accretion in the high-redshift BHs and quasars \cite{suh2025super, ighina2025x}, along with various extreme accretion scenarios \cite{lupi2024sustained}, show support for rapid growth in the early Universe. 
Recent studies of moderate- and low-mass high-redshift AGNs (e.g., Maiolino \emph{et al.} at $4 < z < 11$ (with M$_\text{BH} \sim 10^7 - 10^5 \text{M}_\odot$) \cite{maiolino2024diverse}, Geris \emph{et al.} at $3 < z < 7$ with M$_{\text{BH}} \sim 10^6$M$_\odot$ \cite{geris2026jades}, CAPERS-LRD-z9 AGN at $z = 9.288$ of mass $> 10^7$M$_\odot$\cite{taylor2025capers}, and currently the furthest SMBH in GN-z11 at $z = 10.6$ with mass of $\sim 1.6 \times 10^6$M$_\odot$ \cite{maiolino2024small}) consider accretion-driven growth of light-seed BHs as a viable pathway. This route becomes even more compelling in the presence of the JT-induced TTBR+CE mechanism.

\section{BH-galaxy coevolution and AGN feedback} \label{ssD2ext}

As we previously discussed, AGN outflows will likely accelerate H$_2$ and HD production via the ground and excited manifolds of H$_3^+$. The accelerated cooling by AGN outflows, therefore, may positively regulate both the star formation rates in the host galaxy and the accretion rate of central SMBHs. As a result, AGN feedback acting as a cooling agent could drive the well-established BH-galaxy coevolution, such as the linear positive correlation between the mass of SMBHs, M$_\text{BH}$, and the stellar mass of their host galaxies, M$\star$ \cite{reines2015relations, alexander2025drives}.
Since AGN feedback influences the large-scale structures by processes such as extreme radiation pressure, accretion disc winds, and jets, it has been considered one of the possible candidates for regulating the coevolution, although its complete role remains unresolved \cite{harrison2024observational, alexander2025drives}. 
Observations of AGNs with high accretion rates, often found in gas-rich, star-forming galaxies, further support the role of the feedback \cite{alexander2025drives}.
In the presence of JT-induced TTBR+CE mechanism, AGN-driven H$_2$ production in the host galaxy and the interstellar medium provides a candidate for the physical channel responsible for the BH-galaxy coevolution.
Interestingly, recent JWST observations of low-luminosity and low-mass AGNs/BHs at higher redshift suggest that some early black holes may lie above the local M$_\text{BH}$-M$\star$ relation, potentially indicating rapid early BH growth relative to their host galaxies. \cite{maiolino2024small, maiolino2024diverse, geris2026jades}. For example, AGNs at $4 < z < 11$ (M$_\text{BH} \sim 10^7 - 10^5 \text{M}_\odot$) and at $3 < z < 7$ (M$_{\text{BH}} \sim 10^6$M$_\odot$), reported by Maiolino \emph{et al.} \cite{maiolino2024diverse, maiolino2024small} and Geris \emph{et al.} \cite{geris2026jades}, respectively. The present mechanism also suggests that BH growth in the early Universe would precede the rate of star formation when both are regulated by AGN feedback, aligning with these JWST-detected AGNs at higher redshifts.

In conclusion, we propose a new pathway for molecular hydrogen (H$_2$) and hydrogen deuterium (HD) formation in the early Universe, driven by Jahn-Teller-induced transient three-body recombination and charge exchange in the H$_3^+$ system. Unlike the standard H$^-$ and H$_2^+$ channels, which are strongly suppressed at high redshift by the CMB, this mechanism directly forms ground-state H$_2$ and HD in a single collision. By bypassing fragile intermediates, it may operate efficiently under post-recombination conditions. Enhanced H$_2$ and HD abundances would significantly improve primordial gas cooling, enabling earlier and more efficient fragmentation in low-mass dark matter minihalos, thus shifting the onset of Population III star formation to higher redshifts than predicted by standard models. The resulting acceleration of stellar evolution would provide additional time for the growth of the first black holes, favoring light-seed formation scenarios. The same mechanism in the excited manifolds may also operate in dense, irradiated environments such as AGN outflows, linking molecular chemistry to early black hole accretion and feedback. Overall, this pathway represents a significant extension of primordial chemistry with important implications for early structure formation and recent JWST observations of chemically enriched luminous galaxies and supermassive black holes at high redshift.


\putbib[references-H3+]
\end{bibunit}
\newbibstartnumber{57}
\renewcommand{\refname}{Method References}
\begin{bibunit}[sn-nature]

\renewcommand{\figurename}{Extended Data Fig.}
\renewcommand{\tablename}{Extended Data Table}

\section{Methods} \label{s5meth} 
\subsection{Acronyms and conventions} \label{ssM1abc} 

In the manuscript, whenever it is not stated explicitly, labels $^1\Sigma_g^+$ and $^3\Sigma_u^+$ are used for the ground and first excited electronic states of H$_2$, i.e., $X^1\Sigma_g^+$ and $a^3\Sigma_u^+$ for brevity, respectively. These states are associated with H$(1s)$$+$H$(1s)$ in the asymptotic limit. Similarly, $^2\Sigma_g^+$ and $^2\Sigma_u^+$ are used for the ground and first excited H$_2^+$ states, i.e., $X^2\Sigma_g^+$ and $A^2\Sigma_u^+$, respectively, both associated with H$^+$$+$H$(1s)$ limit. H$_2(g/u)$ and H$_2^+(g/u)$ also represent these states. H and H$^-$ stand for H$(1s)$ and H$^-(1s^2)$, respectively. A similar convention is used for deuterium (D) as well. The main acronyms used in the discussion of JT-induced TTBR+CE mechanism are listed in Extended Data Table \ref{tab:acrnym}.

\setlength{\tabcolsep}{5pt}
\begin{table*}[h]
\centering
\small
\begin{tabular}{|ll|}
\hline
TTBR	&	Transient Three-Body Recombination \\ 
CE		& Charge Exchange \\
JT coupling &	Jahn-Teller coupling \\
ETG	&	Equilateral Triangle Geometry \\
ITG	&	Isosceles Triangle Geometry \\
ADTG & 	Atom-Dimer Triangle Geometry \\
MEC	&	Minimal Energy Configuration \\
PES & Potential Energy Surface \\
\hline
\end{tabular}
\vspace{5pt}
\caption{\justifying $\vert$ \textbf{List of acronyms.} Main acronyms used in the discussion of JT-induced TTBR+CE mechanism for H$_2$ formation.}
\label{tab:acrnym}
\end{table*}

\subsection{JT-induced TTBR+CE reactions on the singlet H$_3^+$ states} \label{ssM1jt} 

\begin{figure*}[t]
\centering
\includegraphics[scale=0.365]{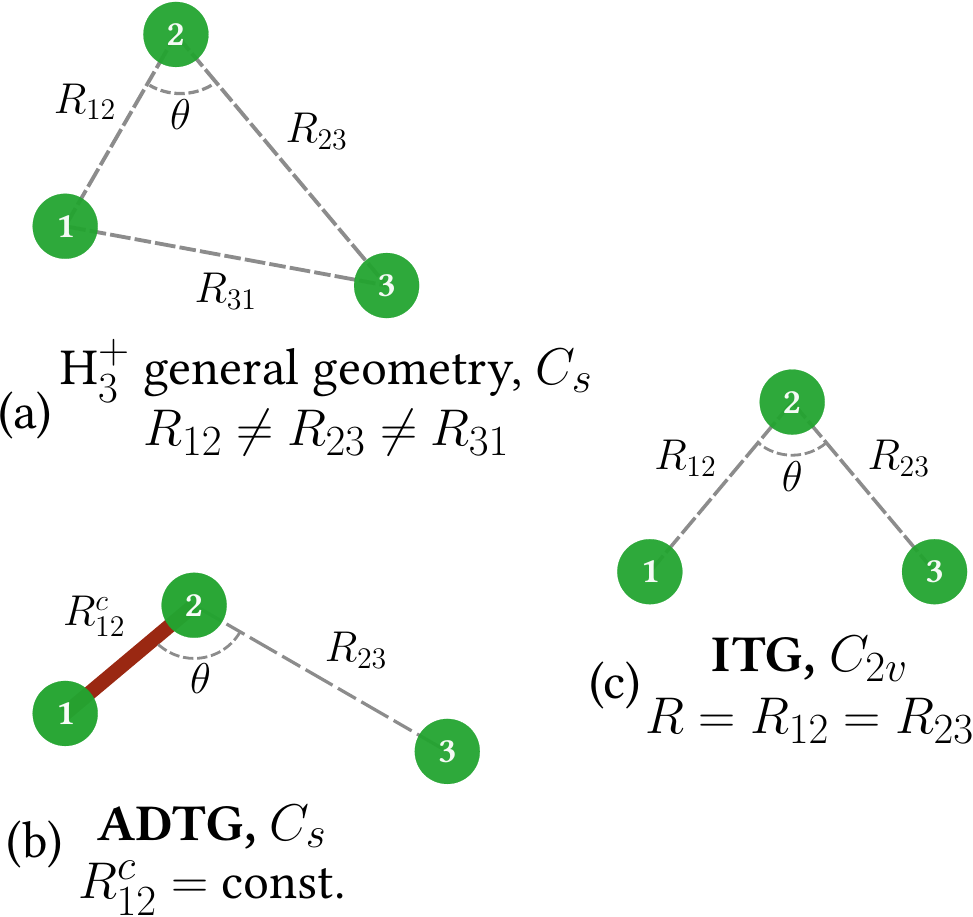}
\caption{\textbf{$\vert$ H$_3^+$ geometries.} (a) Three H$^+$ cores labeled as 1, 2, 3, and the set of internal coordinates $R_{12}$, $R_{23}$, $\theta$ ($\angle$123). (b) The atom-dimer triangle geometry (ADTG), belonging to the $C_{s}$ point group, with $R_{12}$ fixed to a given value $R_{12}^c$. (c) The isosceles triangle geometry (ITG) belonging to the $C_{2v}$ point group, with $R_{12} = R_{23} \equiv R$ for a given $\theta$.}
\label{fig:h2geo}
\end{figure*}

\begin{figure*}[t]
\centering
\includegraphics[scale=0.355]{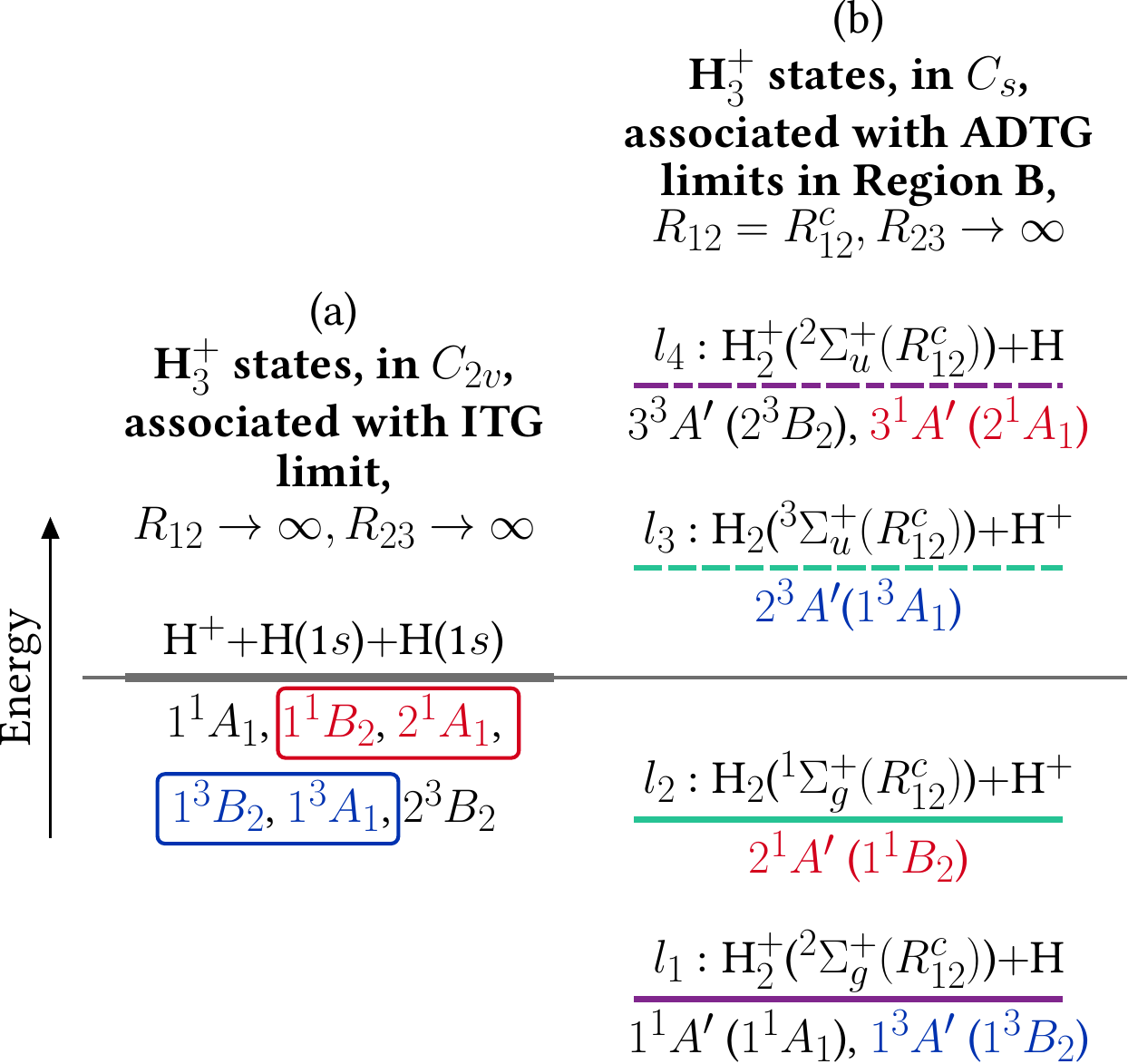}
\caption{$\vert$ \textbf{H$_3^+$ asymptotic limits and states.} (a) For H$^+$, H$(1s)$, H$(1s)$ system in ITG, six H$_3^+$ states are represented in $C_{2v}$ point group and have a common asymptotic limit at $E=0$ cm$^{-1}$, representing $R_{12} = R_{23} \rightarrow \infty$. (b) six H$_3^+$ states in $C_s$ point group, associated with four ADTG asymptotes H$_2 (g/u)$$+$H$^+$(i.e. $l_2$ and $l_3$, resp.) and H$_2^+ (g/u)$$+$H$(1s)$ (i.e. $l_1$ and $l_4$, resp.) for a given dimer size belonging in Region B, ($2.5$a$_0 < R_{12}^c < 10.7$a$_0$, and $l_1 < l_2 < l_3 < l_4$), with $R_{23} \rightarrow \infty$. }
\label{fig:h2asymps}
\end{figure*}

\begin{figure*}[t]
\centering
\includegraphics[scale=0.70]{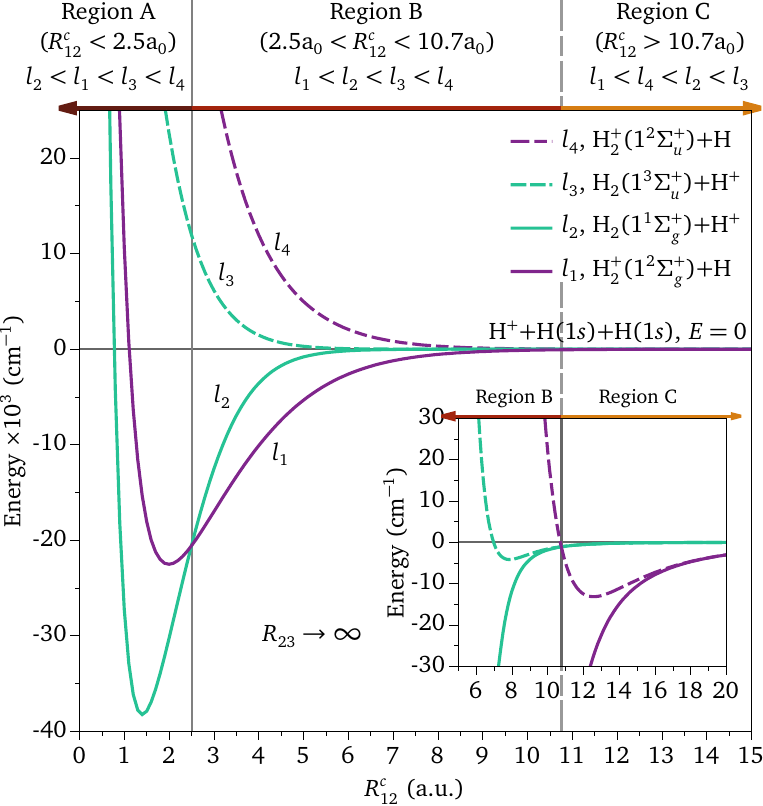}
\caption{$\vert$ \textbf{ADTG asymptotic limits.} Energies of $l_1$, $l_2$, $l_3$, and $l_4$, as a function of the dimer size, $R_{12}^c$, with atom-dimer distance, $R_{23} \rightarrow \infty$. Three Regions, A, B, and C, are identified based on the energy ordering of the ADTG asymptotic limits. The inset shows the onset of Region C where both ADTG asymptotes associated with ionic dimers, H$_2^+ (g/u)$, (i.e., $l_1$ and $l_4$), become lower than the ones associated with neutral ones, H$_2 (g/u)$ , (i.e., $l_2$ and $l_3$). }
\label{fig:h2adtgasymp}
\end{figure*}

To describe the nuclear dynamics of H$^+$, H$(1s)$, and H$(1s)$ system, potential energy surfaces (PESs) of H$_3^+$ is expressed as a function of internal coordinates for three H$^+$ nuclei, namely $R_{12}$, $R_{23}$, and $\theta$($\angle$123), see Extended Data Fig. \ref{fig:h2geo}(a). 
Two special molecular geometries and the respective one-dimensional cuts to the PESs (PECs) are used to address TTBR and CE of Eq. \ref{eq:eqH3ptbr}. 
The first geometry, Atom-Dimer Triangle Geometry (ADTG), keeps the internuclear distance $R_{12}$ fixed at $R_{12}^c$ for a given $\theta$, thereby modeling a non-dissociating dimer within the H$_3^+$ complex, as shown in Extended Data Fig. \ref{fig:h2geo}(b). This setup adequately describes the CE reactions occurring between the charged atom-dimer complex. The second geometry, Isosceles Triangle Geometry (ITG), sets $R_{12} = R_{23} \equiv R$ for a chosen $\theta$, as depicted in Extended Data Fig. \ref{fig:h2geo}(c). This arrangement is suitable for analyzing TTBR in the collision of H$^+$ with two H($1s$) atoms. H$_3^+$ electronic states in ADTG are represented in the low-symmetry $C_s$ point group, exhibiting only a planar symmetry. Whereas H$_3^+$ states in ITG are represented in $C_{2v}$ point group, which has a two-fold rotation axis and two vertical mirror planes. When $\theta = 60$\textdegree, ITG becomes Equilateral Triangle Geometry (ETG), exhibiting additional symmetry elements of $D_{3h}$ point group, which trace the seams of the JT-coupled states.

For H$^+$$+$H($1s$)$+$H($1s$) complex, six states of H$_3^+$ (three singlets and three triplets) in ITG and ADTG are shown in Extended Data Fig. \ref{fig:h2asymps}(a) and (b), respectively, along with their asymptotic limits. For ITG (represented in $C_{2v}$ point group), singlet and triplet H$_3^+$ states are $1^1A_1$, $1^1B_2$(associated with $S1$ of Fig. \ref{fig:JTsch}), $2^1A_1$(associated with $S2$ of Fig. \ref{fig:JTsch}), and $1^3B_2$, $1^3A_1$, $2^3B_2$, respectively (see Extended Data Fig. \ref{fig:h2asymps}(a)). The singlet and triplet JT-coupled pairs are marked by rectangles. In contrast, for ADTG, which is represented in the $C_s$ point group, both singlet and triplet H$_3^+$ states are labeled as $A'$ (see Extended Data Fig. \ref{fig:h2asymps}(b)). The associated ITG states are also shown in round brackets. 
H$_3^+$ states in ITG have a common asymptotic limit with $R_{12} = R_{23} \rightarrow \infty$, representing the fully atomized ion-atom-atom limit, $E = 0$ cm$^{-1}$, shown by a horizontal line in Extended Data Fig. \ref{fig:h2asymps}(a). 
On the other hand, there are four ADTG asymptotic limits, i.e., H$_2^+(g/u)$$+$H ($l_1$ and $l_4$) and H$_2(g/u)$$+$H$^+$ ($l_2$ and $l_3$), as shown in Extended Data Fig. \ref{fig:h2asymps}(b). It is important to notice that a singlet and a triplet H$_3^+$ states are associated with both ADTG limits with ionic dimers, i.e., $l_1$ and $l_4$. For ADTG limits with neutral dimers, $l_2$ and $l_3$, only one state (either singlet or triplet) is connected to each limit.

Energies of the four ADTG asymptotic limits, $l_1$, $l_2$, $l_3$, and $l_4$, as a function of the dimer size, $R_{12}^c$, for $R_{23} \rightarrow \infty$, are shown in Extended Data Fig. \ref{fig:h2adtgasymp}. Region A ($R_{12}^c < 2.5$a$_0$, with $l_2 < l_1 < l_3 < l_4$), Region B ($2.5$a$_0 < R_{12}^c < 10.7$a$_0$, with $l_1 < l_2 < l_3 < l_4$), and Region C ($R_{12}^c > 10.7$a$_0$, with $l_1 < l_4 < l_2 < l_3$) mark the domains in which the energy ordering of the four ADTG limits are different.

H$_3^+$ PESs are computed under the Born-Oppenheimer approximation using the MOLPRO package \cite{MOLPRO-WIREs}. The complete active-space self-consistent field (CASSCF) and multireference configuration interaction (MRCI-SD) methods, which include single and double electronic excitations, are utilized \cite{werner1988efficient}. Calculations use an augmented correlation-consistent basis set, av5z, for hydrogen atoms and are benchmarked against previous studies \cite{viegas2007accurate, aguado2021three}.

\begin{figure*}[t]
\centering
\includegraphics[scale=0.205]{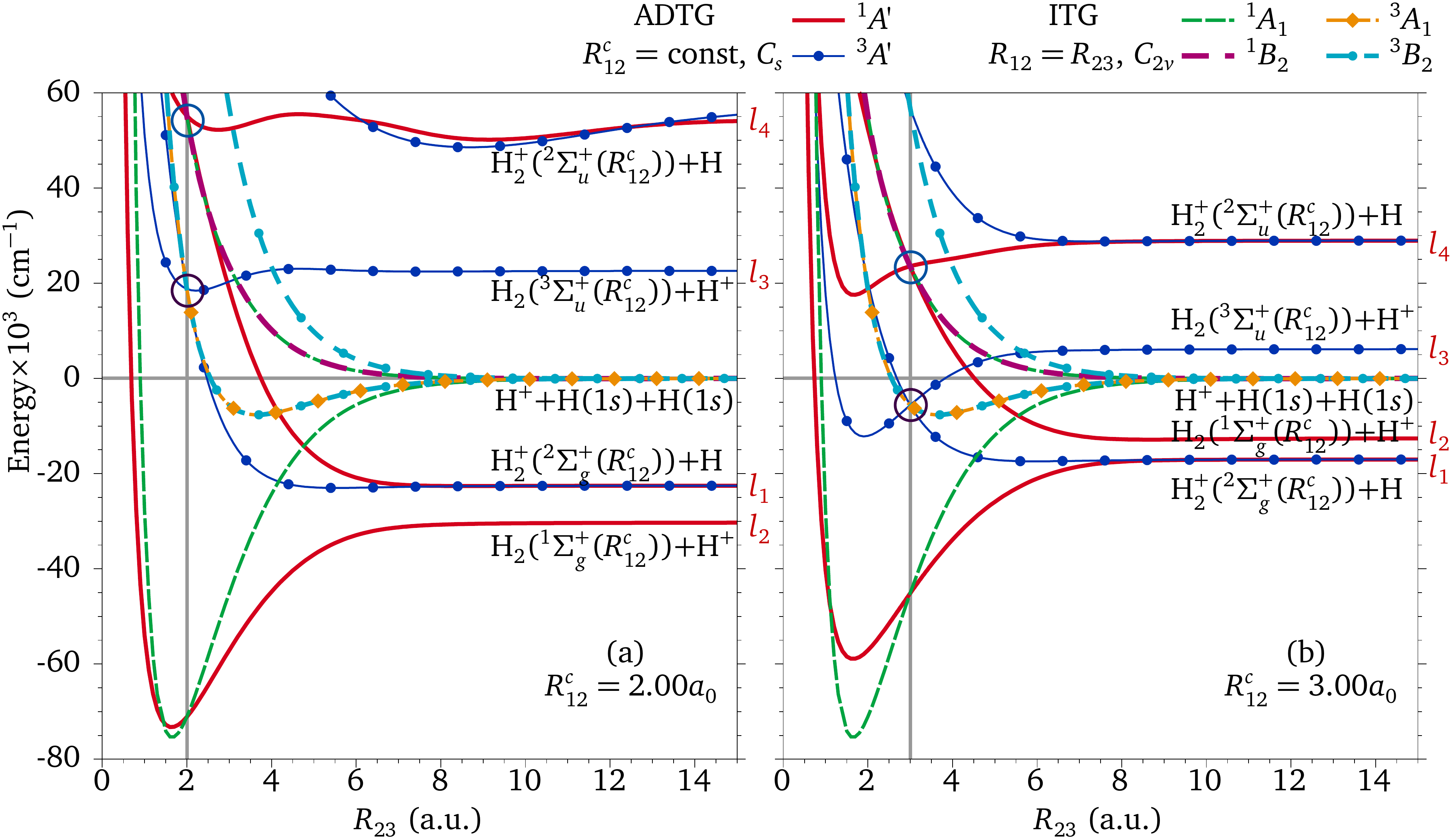} 
\caption{$\vert$ \textbf{H$_3^+$ states in ADTG and ITG.} Three singlet and three triplet states of H$_3^+$ in ADTG and ITG for $\theta=60$\textdegree~associated with the first fully atomized limit, H$^+$$+$H($1s$)$+$H($1s$). ADTG states are presented for $R_{12}^c = 2.00 a_0$ (corresponds to Region A, Extended Data Fig. \ref{fig:h2adtgasymp}), and for $R_{12}^c = 3.00 a_0$ (Region B, Extended Data Fig. \ref{fig:h2adtgasymp}), in Extended Data Fig. \ref{fig:h3pecs} (a) and (b), respectively. The energy ordering change of the ADTG asymptotes ($l_1$, $l_2$, $l_3$, $l_4$) is also indicated. ITG H$_3^+$ states, which are identical in both panels, all dissociate to H$^+$$+$H($1s$)$+$H($1s$). Vertical lines at $R_{23} = 2.00 a_0$ in Extended Data Fig. \ref{fig:h3pecs}(a) and $R_{23} = 3.00 a_0$ in Extended Data Fig. \ref{fig:h3pecs}(b) mark the geometries where the JT-coupling condition is met, i.e., $R_{23} = R_{12}^c $, $\theta = 60$\textdegree~(the ETG). Circles highlight observed JT couplings in the second and third singlet, $2^1A'$ \& $3^1A'$; $S1$ and $S2$ of Fig. \ref{fig:JTsch}, and first and second triplet, $1^3A'$ \& $2^3A'$, H$_3^+$ states. The corresponding singlet ITG states, $1^1B_2$, $2^1A_1$, that are degenerate at $\theta = 60$\textdegree, represent the seam of the JT-coupling in $S1$ and $S2$, shown in Fig. \ref{fig:JTsch}(d). For the triplet JT-coupled pairs, these are $1^3B_2$, $1^3A_1$ states.}
\label{fig:h3pecs}
\end{figure*}

In Extended Data Fig. \ref{fig:h3pecs} (a) and (b), ADTG and ITG states of H$_3^+$ are shown for $\theta=60$\textdegree, in the internuclear separation space, where both $R_{12}$ and $R_{23}$ are finite. For the ADTG states, $R^c_{12}$ is $ 2.00 a_0$ in Extended Data Fig. \ref{fig:h3pecs}(a), representative of Region A. In Extended Data Fig. \ref{fig:h3pecs}(b), $R^c_{12}$ is $3.00 a_0$,  representative of Region B. The energy orderings of the ADTG asymptotes are provided in both panels; as we can see, they are consistent with Extended Data Fig. \ref{fig:h2adtgasymp}. Vertical lines at $R_{23} = 2.00 a_0$ in Extended Data Fig. \ref{fig:h3pecs}(a) and $R_{23} = 3.00 a_0$ in Extended Data Fig. \ref{fig:h3pecs}(b) mark the evolution of ADTG states to the ETG, where the condition for the JT-coupling is met, i.e., $R_{23} = R_{12}^c $, $\theta = 60$\textdegree. At these points, marked by circles above the vertical lines, the second and third singlet states ($2^1A'$, $3^1A'$; $S1$ and $S2$ from Fig. \ref{fig:JTsch}) and the first and second triplet states ($1^3A'$, $2^3A'$) become degenerate. The corresponding ITG singlet ($1^1B_2$, $2^1A_1$; associated with $S1$ and $S2$ of Fig. \ref{fig:JTsch}(d)) and triplet ($1^3B_2$, $1^3A_1$) states are degenerate for all internuclear separations at ETG and represent the seams of JT coupling in H$_3^+$. For $\theta \neq 60$\textdegree, ADTGs represent isosceles triangles at $R_{23} = R_{12}^c$, which lifts the degeneracy between JT-coupled states. The influence of geometry on JT-induced TBR and CE collisions in a similar system, Rb$_3^+$, has been discussed previously \cite{pandey2024ultracold}.

\begin{figure*}[t]
\centering
\includegraphics[scale=0.345]{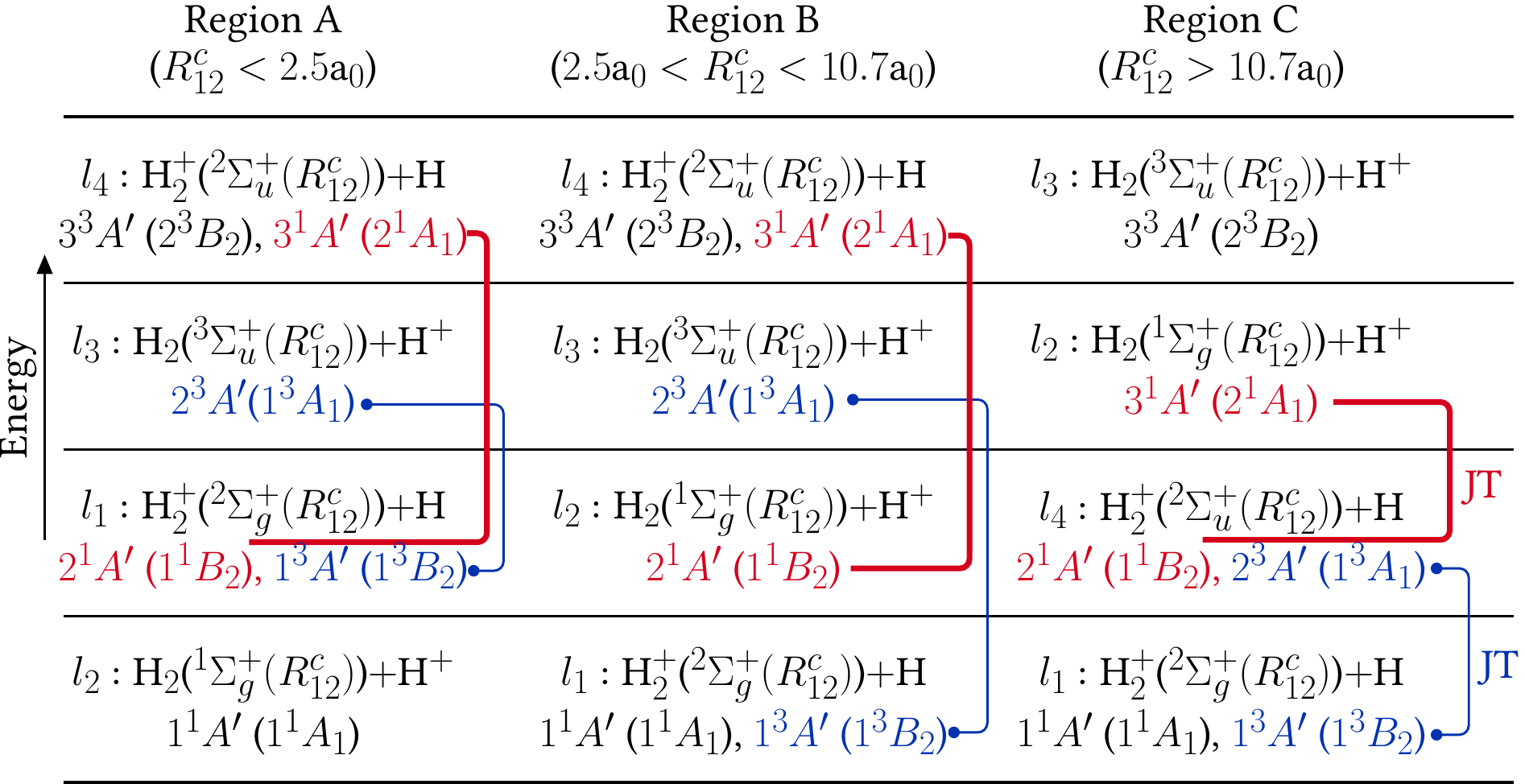}
\caption{$\vert$ \textbf{Energy ordering of ADTG limits.} Energy ordering of $l_1$, $l_2$, $l_3$, $l_4$, in Regions A, B, and C. The JT-coupled ADTG singlet ($2^1A' (S1)$ \& $3^1A' (S2)$) and triplet ($1^3A'$ \& $2^3A'$) H$_3^+$ electronic states, linked by the right brackets, are associated with different ADTG limits in Regions A, B, and C. The possibility of CE in the JT-coupled singlet states, $S1$ and $S2$, involving the transition state \{H$_2^+(^2\Sigma_u^+) +$H$(1s)$\} of Eq. \ref{eq:eqH3ptbr}, is decided by whether the outgoing JT-coupled state leads to H$_2 (g)$ dimer in the ADTG limit (as depicted in Fig. \ref{fig:JTsch}(e)).}
\label{fig:JTregions}
\end{figure*}

The CE channel in Eq. \ref{eq:eqH3ptbr} can proceed via JT-induced coupling between the singlet H$_3^+$ states, $S1$ and $S2$. 
Whether the JT coupling permits CE reactions is determined by the energy ordering of the ADTG asymptotes in Regions A, B, and C, as shown in Extended Data Fig. \ref{fig:JTregions}. Specifically, the constraint arises because the JT coupling consistently links the second and third singlet states, $2^1A'$($S1$) and $3^1A'$($S2$), across all internuclear separations. 
For Eq. \ref{eq:eqH3ptbr}, the transition state \{H$_2^+(^2\Sigma_u^+) +$H$(1s)$\} is associated with $2^1A'$($S1$) in Region C and with $3^1A'$($S2$) in Region B. 
In both Regions B and C, this atom-dimer transition complex then connects, via Jahn-Teller coupling, to H$_2(^1\Sigma_g^+) +$H$^+$, enabling charge exchange.
In contrast, in Region A, as shown in Extended Data Fig. \ref{fig:JTregions}, the JT-coupled singlet state $3^1A'$, associated with \{H$_2^+(^2\Sigma_u^+) +$H$(1s)$\}, does not support charge exchange, but only electronic symmetry exchange in H$_2^+$, i.e., $u \leftrightarrow g$.

\begin{figure}[t]
\centering
\includegraphics[scale=0.415]{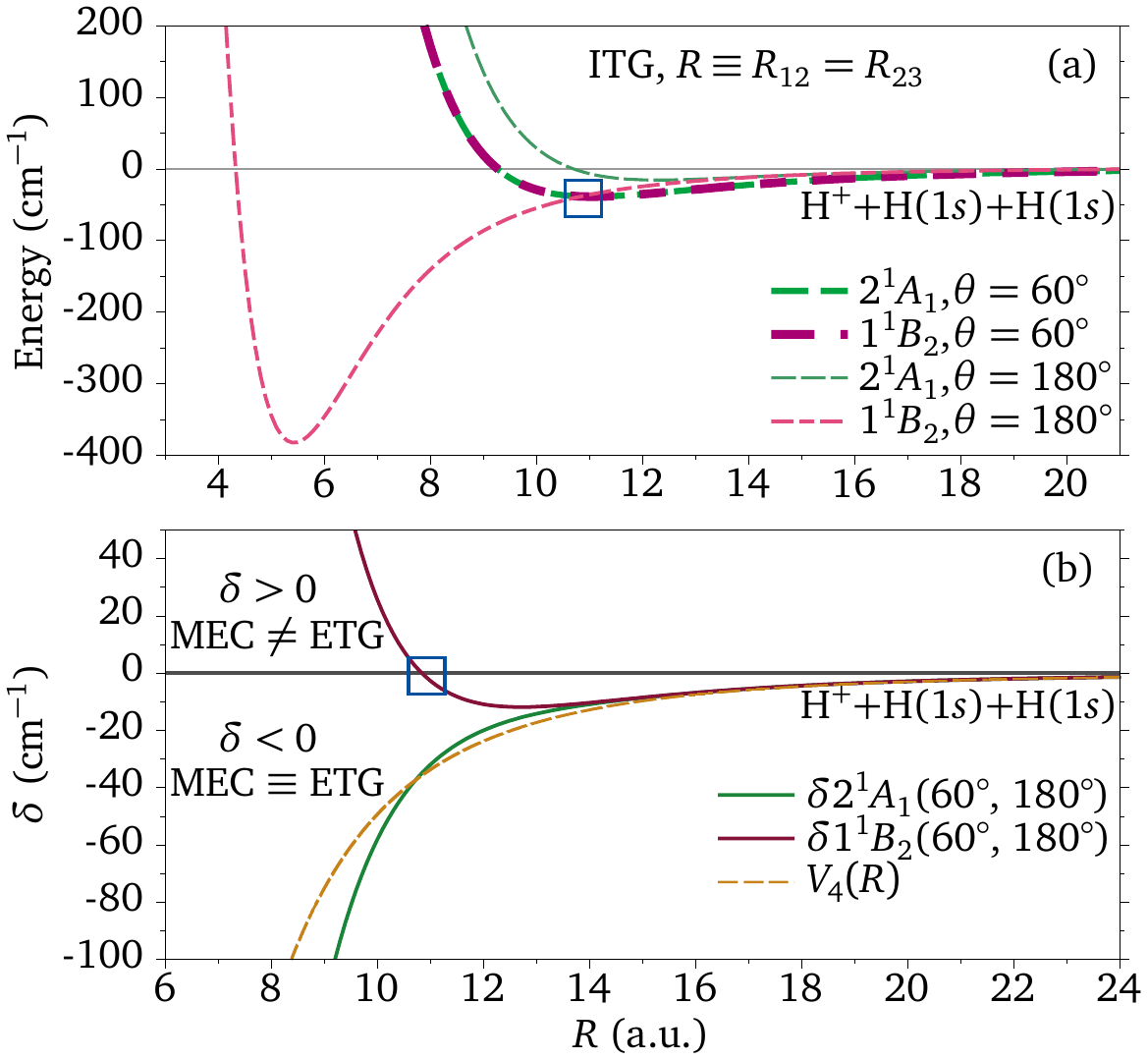}
\caption{$\vert$ \textbf{Minimal energy configuration determination.} (a) One-dimensional cuts of the PESs of H$_3^+$ ITG states $1^1B_2$, $2^1A_1$ (associated with ADTG $2^1A'$($S1$) \& $3^1A'$($S2$) states, respectively) involved in the JT coupling, for $R_{12}=R_{23} \equiv R <21a_0$, and for $\theta=60$\textdegree~and 180\textdegree. Both curves are degenerate for 60\textdegree~(the ETG, identical to one depicted in Fig. \ref{fig:JTsch}(d)) with the well depth of $-38.7$ cm$^{-1}$ at $R = 11.10 a_0$. Panel (b) shows the energy difference, $\delta$, between the PECs at $\theta=60$\textdegree~and 180\textdegree, highlighting that for the $1^1B_2$ curve, the latter geometry has a minimal energy for $R < 10.9 a_0$ (squared box). For the large $R$, $\delta$ for both states show $V_4(R)$ nature, where $V_4(R) = -C{_4}/R{^4}$, with $2C{_4} = \alpha_d = 4.5$ a.u. $2^1A_1$ shows its minima for 60\textdegree~geometry for all molecular sizes.}
\label{fig:H3c2vsmin}
\end{figure}

Another important constraint in the TTBR and CE reactions is that JT-coupling occurs only at equilateral triangles, as shown in Fig. \ref{fig:JTsch}(d). For a collision between H$^+$, H$(1s)$, and H$(1s)$ that begins in a random orientation in hydrogen gas, the equilateral triangle must be the MEC. If so, geometry rearrangement during the collision can then drive the system toward the ETGs. 
To describe the geometry rearrangement on the JT-coupled singlet states of H$_3^+$, their ITG states, 1$^1B_2$ and 2$^1A_1$ (associated with $S1$ and $S2$, respectively), are shown for $\theta=60$\textdegree~and $\theta=180$\textdegree~in Extended Data Fig. \ref{fig:H3c2vsmin} (a). Energies of all other geometries for 60\textdegree $< \theta < 180$\textdegree~fall in between these two extremes. As discussed earlier, 1$^1B_2$ and 2$^1A_1$ ITG curves are degenerate for $\theta=60$\textdegree~at all internuclear separations, representing the seam of the JT-coupled states ( also shown in Extended Data Fig. \ref{fig:h3pecs} (a) and (b)). The energy differences between these two geometries for the states, $\delta 1^1\text{B}_2$(60\textdegree, 180\textdegree) and $\delta 2^1\text{A}_1$(60\textdegree, 180\textdegree), are shown in Extended Data Fig. \ref{fig:H3c2vsmin} (b). A negative $\delta$ represents the minima for $\theta = 60$\textdegree, i.e.,  the region where MEC is identical to the ETG, enabling the occurrence of JT-coupling. When $\delta > 0$, linear geometry ($\theta = 180$\textdegree) becomes the MEC, thereby prohibiting the JT-coupling.  
In Region B, the transition complex \{H$_2^+(^2\Sigma_u^+) +$H$(1s)$\} of Eq. \ref{eq:eqH3ptbr} is associated with the 2$^1A_1$ ITG state (see Extended Data Fig. \ref{fig:JTregions}). This state exhibits minima for $\theta=60$\textdegree~for all interparticle separations, i.e., $\delta < 0~\forall ~R$. Thus, ensuring the occurrence of Jahn-Teller coupling between H$_3^+$ singlet states. In Region C, on the other hand, \{H$_2^+(^2\Sigma_u^+) +$H($1s$)\} is associated with the 1$^1B_2$ ITG state. This state shows minima for the ETG for $R (\equiv R_{12} = R_{23}) > 10.9 a_0$, i.e., $\delta < 0~\forall ~R > 10.9 a_0$, marked by a square in Extended Data Fig. \ref{fig:H3c2vsmin} (a) and (b). It covers almost the entire Region C, defined in the ADTG asymptotic limit as $R_{12}^c > 10.7 a_0$ with $R_{23} \rightarrow \infty$. Therefore, in Region C, as well, TTBR and CE would occur. These possibilities are summarized in Extended Data Table \ref{tab:h3pgeo}.

\setlength{\tabcolsep}{6pt}
\begin{table*}[t!]
\centering
\small
\begin{tabular}{|l|l|l|}
\hline
&  &\\ 
\multicolumn{1}{|c|}{ADTG states of} & 
\multicolumn{1}{c|}{JT-coupled outgoing} &
\multicolumn{1}{c|}{\multirow{2}{*}{CE}} \\
\multicolumn{1}{|c|}{\{H$_2^+$($^2\Sigma_u^+$($R^c_{12}$))$+$H\} in the asymptotic limit; $R_{23} \rightarrow \infty$} & 
\multicolumn{1}{c|}{channels and H$_3^+$ states} &
\multicolumn{1}{c|}{} \\
&  &\\ 
\hline
Region A, $R^c_{12} < 2.5\text{a}_0$,\hspace{3pt}  $3^1A'$(60\textdegree~$\forall$ $R$)  & H$_2^+$($X^2\Sigma_g^+$)$+$H,\hspace{7pt}  $2^1A'$ & No \\
Region B, $2.5\text{a}_0 < R^c_{12} < 10.7\text{a}_0$,\hspace{3pt}  $3^1A'$(60\textdegree~$\forall$ $R$)  & H$_2$~($X^1\Sigma_g^+$)$+$H$^+$,   $2^1A'$ & Yes \\
Region C, $R^c_{12} > 10.7\text{a}_0$,\hspace{3pt} $2^1A'$(60\textdegree~$\forall$ $R > 10.9 a_0$) & H$_2$~($X^1\Sigma_g^+$)$+$H$^+$,  $3^1A'$ & Yes \\
\hline
\end{tabular}
\vspace{5pt}
\caption{\justifying $\vert$ \textbf{Charge exchange prospects.} The possibilities of charge exchange between H$_2^+$($^2\Sigma_u^+$($R^c_{12}$)) and H, i.e., the transition state of Eq. \ref{eq:eqH3ptbr}, in the ADTG asymptotic limit in Regions A, B, and C are summarized, combining the information from Extended Data Fig. \ref{fig:h2adtgasymp}, \ref{fig:JTregions}, and \ref{fig:H3c2vsmin}. The initial ADTG electronic state with the position in $\theta$ of its minimal energy configuration for ITG is also provided in the first column, recalling that CE will be possible via JT coupling only for $\theta = 60$\textdegree. In the second column, ADTG asymptotes and the associated states of the outgoing channels are provided. In Regions B and C, CE is allowed via JT coupling. However, in Region A, JT-induced reaction would only allow exchange of H$_2^+$ molecular ion symmetry from $u$ to $g$.}
\label{tab:h3pgeo}
\end{table*}

It is important to note that, in the current manuscript, we left out two reaction channels. The first involves reactions on the triplet JT-coupled states of H$_3^+$. Collisions on the JT-coupled triplet states would produce H$_2^+(^2\Sigma_g^+)$ in a TBR reaction. However, the subsequent CE channel would connect them to the ungerade states of either H$_2$ or H$_2^+$, which are either weakly bound or unstable (see Extended Data Fig. \ref{fig:JTregions}). Eq. \ref{eq:eqH3ptbr} also leaves out the TBR collisions that directly create neutral H$_2$. This is because the TBR rate for producing weakly bound neutral dimers would be much lower---about 2 to 4 orders of magnitude---than that for producing weakly bound molecular ions, as discussed for many ion-atom-atom cases \cite{cretu2022ion, mirahmadi2023ion}.

\subsection{Comparison of H$^-$ and H$_2^+$ mechanisms with JT-induced TTBR+CE mechanism at high-redshifts} \label{ssD3supp}

The standard pathways of H$_2$ formation involving H$^-$ and H$_2^+$ intermediates, and their role in the creation of first stars in the early Universe were identified as early as the 1960s \cite{saslaw1967molecular, peebles1968origin}. Saslaw \& Zipoy first calculated H$_2$ abundance based on the H$_2^+$ mechanism in the primordial Universe \cite{saslaw1967molecular}. H$_2$ production by the H$^-$ mechanism was first suggested by Peebles \& Dicke \cite{peebles1968origin}. 
In the H$^-$ mechanism, H$_2$ formation occurs by a two-step reaction pathway involving the H$^-$ ion created by radiative attachment as an intermediate step \cite{mcdowell1961formation, peebles1968origin}, 

\begin{equation}
\text{H}(1s) + \text{e}^- \rightarrow \text{H} ^-(1s^2) + \gamma,
\label{eq:eqH2a1}
\end{equation}

followed by production of H$_2$ by associative detachment,

\begin{equation}
\text{H}^-(1s^2) + \text{H}(1s) \rightarrow \text{H}_2(^1\Sigma_g^+, v_g) + \text{e}^-.
\label{eq:eqH2a2}
\end{equation}

H$_2$($^1\Sigma_g^+, v_g$) produced by Eq. \ref{eq:eqH2a2} are predominantly created in its excited vibrational levels $v_g = 4 - 7$ \cite{bieniek1979associative}. As discussed earlier, the H$^-$ mechanism is the dominant reaction channel for H$_2$ formation at redshift $z \simeq 30 - 20$ when the first stars formed \cite{hirata2006cosmological}. 
Second pathway of H$_2$ formation, referred as H$_2^+$ mechanism, involves production of H$_2^+$ by radiative association \cite{saslaw1967molecular},

\begin{equation}
\text{H}(1s) + \text{H}^+ \rightarrow \text{H}_2^+(^2\Sigma_g^+, v_g^+) + \gamma,
\label{eq:eqH2b1}
\end{equation}

followed by the charge transfer reaction,

\begin{equation}
\text{H}_2^+(^2\Sigma_g^+, v_g^+) + \text{H}(1s) \rightarrow \text{H}_2(^1\Sigma_g^+, v_g) + \text{H}^+.
\label{eq:eqH2b2}
\end{equation}

\setlength{\tabcolsep}{1pt}
\begin{table*}[t]
\centering
\small
\begin{tabular}{|l|c|c|c|}
\hline
& \multirow{2}{*}{H$^-$ mechanism} & \multirow{2}{*}{H$_2^+$ mechanism} & \multirow{2}{*}{JT-induced TTBR+CE}\\
& & & \multirow{2}{*}{mechanism} \\
& & & \\
\hline
Production & Two-step  & Two-step & One-step \\
processes & Intermediate: H$^-$ & Intermediate: H$_2^+$ & No intermediate \\
\hline 
Suppression & Inverse of the 1st-step. & Inverse of the 1st-step. & \multirow{2}{*}{Not affected by CMBR} \\
 processes & CMBR. & CMBR. & \\
\hline
Primary source &  at $z < 100$ & at $100 < z < 300$ & at $z \leq 1100$, AGN outflows \\ 
\hline
main HD & Secondary & Secondary & \multirow{2}{*}{Direct} \\
production & bond-rearrangement & bond-rearrangement &  \\
\hline
Exited state & \multirow{2}{*}{No excited state} & Strongly suppressed & Multiple pathways in H$_3^{+*}$,\\
channels & & (if present) & possibly active in AGNs\\ 
\hline
\multicolumn{1}{|l}{Formation of} & 
\multicolumn{2}{|c|}{\multirow{2}{*}{at $20 < z < 30$}} &
\multicolumn{1}{c|}{\multirow{2}{*}{at $30 < z < 50$ (see text)}} \\
\multicolumn{1}{|l}{the first stars} & 
\multicolumn{2}{|c|}{\multirow{2}{*}{Pop III.1 pathway simulation}} &
\multicolumn{1}{c|}{\multirow{2}{*}{HD-dominated cooling}} \\
\multicolumn{1}{|l}{and black holes} & 
\multicolumn{2}{|c|}{} &
\multicolumn{1}{c|}{} \\
\hline
\end{tabular}
\vspace{5pt}
\caption{\justifying $\vert$ \textbf{Main H$_2$ formation pathways.} A comparison of the standard H$^-$ and H$_2^+$ mechanisms for H$_2$ formation with the proposed JT-induced TTBR+CE mechanism. The major advantage of JT-induced TTBR+CE mechanism, being a single-step reaction, is its immunity against photodissociation by CMB radiation, which strongly suppresses the H$^-$ and H$_2^+$ intermediates and limits H$_2$ production in the early Universe ($z > 100$) (see text for details). Given this, H$_2$ and HD could be produced from the JT-induced TTBR+CE mechanism much earlier, right towards the end of the epoch of recombination ($z \leq 1100$), which may facilitate the efficient molecular cooling. In the Pop III.2 scenario, HD-dominated cooling occurs only at later stages, when HD production rises from the bond-rearrangement reaction, a secondary process dependent on high abundances of H$_2$ and D$^+$, and therefore requiring additional favorable conditions. 
In contrast, HD formation via the JT-induced TTBR+CE mechanism could enable early star formation in HD-cooled primordial gas. Depending on the reaction rate, the first stars and black holes might appear significantly earlier. Additionally, JT-coupled reactions producing H$_2$ and HD in the excited states of H$_3^{+*}$ could provide previously overlooked cooling channels in AGNs and their outflows, enhancing star formation and black hole growth, and driving BH-galaxy coevolution.
}
\label{tab:h2mech}
\end{table*}

Radiative association of H$(1s)$ and H$^+$ creates H$_2^+$($^2\Sigma_g^+, v_g^+$) mostly in its excited vibrational levels $v_g^+ > 10$ \cite{sugimura2016role}. 
The charge exchange reaction H$_2^+$ $^2\Sigma_g^+$ $(v_g^+,j^+)$ $+$ H $\rightarrow$ H$_2$$^1\Sigma_g^+$($v_g,j$) $+$ H$^+$ has been studied extensively. 
These studies focus on calculating cross-sections and exploring vibrational level dependence \cite{ghosh2017beyond, mukherjee2019beyond, ghosh2021charge, sanz2021near, roncero2022vibrational, aguado2021three, krstic2002inelastic, krstic2005vibrationally}. 
It is important to note that in charge exchange reactions, such as the one given in Eq. \ref{eq:eqH2b2}, and in the three-body, diatomic association like H$^+$$+$H$(1s)$$+$H$(1s)$, studied by Krstic \emph{et al.} \cite{krstic2003three}, H$_2^+$ and H$_2$ are both considered in their ground states, i.e., H$_2^+$($X^2\Sigma_g^+$) and H$_2$($X^1\Sigma_g^+$). 
Their charge-exchange interactions with H$(1s)$ or H$^+$ are therefore described by couplings between the first and second singlet states of the H$_3^+$ complex, rather than by the Jahn-Teller-coupled second and third singlet states \cite{ghosh2017beyond, mukherjee2019beyond, ghosh2021charge, sanz2021near, roncero2022vibrational, aguado2021three, krstic2002inelastic, krstic2005vibrationally, krstic2003three}.

The initial steps of both the H$^-$ and H$_2^+$ formation mechanisms (Eq. \ref{eq:eqH2a1} and Eq. \ref{eq:eqH2b1}, respectively) are intrinsically slow \cite{faure2024chemistry}. Nevertheless, the  H$^-$ pathway is generally more efficient under typical conditions, as its formation rate is roughly an order of magnitude higher than that of H$_2^+$ \cite{faure2024chemistry, sugimura2016role}. At high redshift, however, the CMB strongly suppresses this channel. 
H$^- (1s^2)$ is only weakly bound ($\sim -$0.75 eV \cite{lykke1991threshold}, whereas H$_2^+$($^2\Sigma_g^+,v_g^+=0$) is more strongly bound ($\sim -$2.65 eV \cite{zammit2017state})), making photodetachment particularly effective for H$^-$. Consequently, for $z > 100$ the H$^-$ route contributes very little to H$_2$ production, and the H$_2^+$ pathway becomes the dominant formation channel \cite{galli1998chemistry, sugimura2016role}.

Even so, H$_2^+$ formed by radiative association is mainly created in highly excited vibrational levels of the $X^2\Sigma_g^+(v_g^+)$ state. Decay to lower vibrational levels is slow, occurring only via weak quadrupole transitions. In contrast, as photodissociation back to the $A^2\Sigma_u^+(v_u^+)$ continuum is a dipole-allowed transition, most newly formed H$_2^+$ is destroyed before it reacts to form H$_2$. This vibrational sensitivity also makes the H$_2^+$ channel inefficient at early times, further lowering the primordial H$_2$ abundance \cite{hirata2006cosmological, sugimura2016role}.
Aspects of H$_2$ formation from the H$^-$ and H$_2^+$ mechanisms, along with the proposed JT-induced TTBR+CE mechanism, are summarized in Extended Data Table \ref{tab:h2mech}.

\subsection{Energy constraint for JT-induced reactions at the high-redshift thermodynamic conditions} \label{ssM3thrm}  

We previously found that JT coupling between the $S1$ and $S2$ states of H$_3^+$ occurs only at ETGs. On the other hand , three-body collisions involving H$^+$ and two H($1s$) atoms begin from random geometries, with their evolution determined by the interplay between the total initial kinetic energies (KEs) of the particles and the system's interaction potential, which drives it toward the minimum energy configuration (MEC). The initial KEs of H$^+$ and the H($1s$) atoms are depicted in Fig. \ref{fig:JTsch}(a). It is important to note that the nuclei's KEs are not included in the calculation of the potential energy surfaces (PESs) under the fixed-nuclei Born-Oppenheimer approximation. Instead, nuclear KEs influence the collision dynamics, with the computed PES serving as the guiding potential for their motion.

The well depth of H$_3^+$ singlet states, $S1$ and $S2$ ($1^1B_2$ and $2^1A_1$, respectively, in ITG; both are degenerate at ETGs), is $-38.7$ cm$^{-1}$ for an equilateral triangle of sides equal to 11.10a$_0$. See Extended Data Fig. \ref{fig:H3c2vsmin} (a). In addition to the condition for JT-induced reactions shown in Fig. \ref{fig:JTsch}(d), the finite well depth of the PESs adds further restrictions on the reaction. Specifically, the sum of the initial KEs of the ion-atom-atom triad must be less than the well depths of $S1$ and $S2$ for the participants to be effectively guided by these PESs. To assess the effect of this restriction, we consider a simplified picture. In this model, each colliding partner—the H$^+$ ion or each of the two H atoms—should have KE below $12.9$ cm$^{-1}$ (meaning $3 \times 12.9~\text{cm}^{-1} < 38.7$ cm$^{-1}$) for geometry rearrangement to occur during collisions. In the present analysis, we have assumed a rather stronger condition on particles' energies, KE $< 12$ cm$^{-1}$.

\begin{figure}[t]
\centering
\includegraphics[scale=0.595]{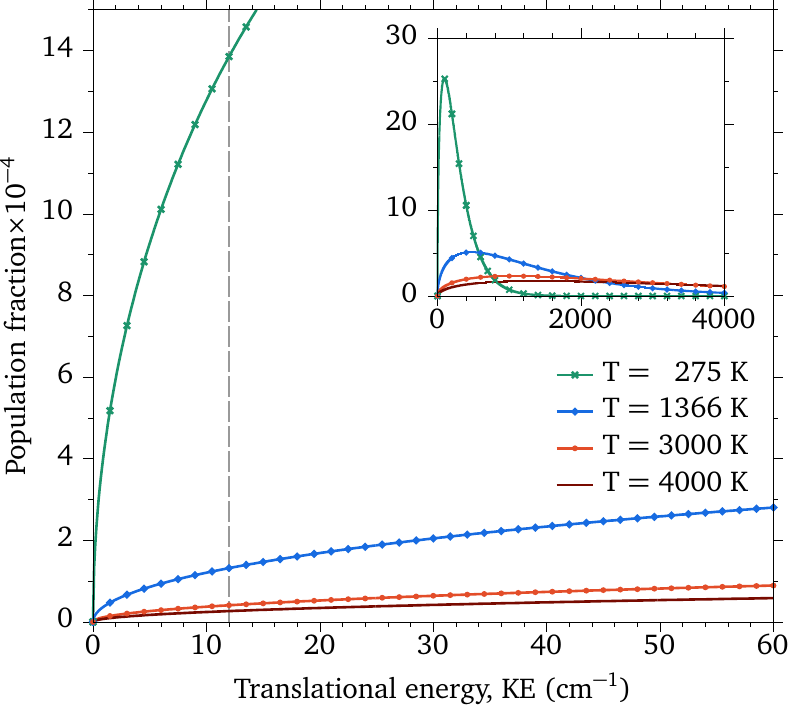}
\caption{$\vert$ \textbf{Energy constrained population fraction.} Maxwell-Boltzmann probability distributions for T $=$ 275 K, 1366 K, 3000 K, and 4000 K, representing the CMB temperatures at $z=100, 500, 1100$, and 1500, respectively. The vertical line at $12$ cm$^{-1}$ marks the kinetic energy limit, (KE $< 12$ cm$^{-1}$ ), for H or H$^+$ in the ion-atom-atom collision, relevant to Jahn-Teller reactions on singlet H$_3^+$ PESs, $2^1A'$($S1$) and $3^1A'$($S2$), needed for geometry rearrangement. The inset shows the probability distributions over a wider energy range.}
\label{fig:MB-energy}
\end{figure}

If the well depths of the PESs imposed no energy constraints, the entire H gas ensemble at temperature T would be available for such a reaction. However, because the present case requires particle KE to be $<$ 12 cm$^{-1}$, at any given moment only a fraction, P$_k$, of the thermal gas has H atoms with KE $<$ 12 cm$^{-1}$. For T = 275 K, 1366 K, 3000 K, and 4000 K, P$_k$(KE $<$ 12 cm$^{-1}$) are 0.011363, 0.001059, 0.000327, and 0.000213, respectively, as shown in Extended Data Fig. \ref{fig:MB-energy}. 
Meanwhile, due to ongoing collisions among H atoms, the subset of atoms satisfying the KE $<$ 12 cm$^{-1}$ constraint is constantly changing. As a result, over time, every H atom in the gas can meet the energy constraint at some point. 
These collisions also ensure that whenever low-energy H atoms are removed through TTBR and CE reactions, other H atoms adjust to satisfy the KE $<$ 12 cm$^{-1}$ constraint. This process maintains the Maxwell-Boltzmann distribution, preserving P$_k$ as a statistical invariant.

For the JT-induced reaction to remain unaffected by the energy constraint, the timescale in which the entire gas ensemble satisfies the energy constraint must be much shorter than the reaction timescales. 
The time required for the full hydrogen gas ensemble to reach a state in which the kinetic energy of each H atom satisfies KE $<$ 12 cm$^{-1}$ at least once is $t_R =$ (1/P$_k$)$\star$$t_f$.
Where the mean-free time, $t_f$, for hard-sphere elastic collisions that change the speed of H atoms is $t_f = 1/(\sigma \times <v> \times n_\text{H})$; $\sigma = \pi (2 r)^2$ is the collision cross section with H atoms of radius $r = 2.5$ a$_0$, $<v>$ is the mean speed of H atoms at temperature T, and $n_\text{H}$ is the hydrogen gas density. 
For T $=$ 3000 K ( representative of CMB temp. at $z=1100$ with $n_\text{H} \sim 2.1\times 10^3$ cm$^{-3}$) and T $=$ 4000 K ($z=1500$ with $n_\text{H} \sim 5.4\times 10^3$ cm$^{-3}$), $t_R$ are 74.42 and 45.15 years, respectively. Varying the hard-sphere size in the range of $r = 2.0 \text{a}_0$ to $4.0 \text{a}_0$ that covers the strong interaction domain for H$_2$$X^1\Sigma_g^+$ state would change the $t_R$ by factors of 1.56 to 0.40. In short, at high redshifts near the epoch of recombination, $t_\text{R}$ would be a few tens of years.

Using the probability P$_k$ at temperature T, we also estimate the time, $t_N$, for three-body collisions where each participant must have KE less than 12 cm$^{-1}$. In $t_N$ time, every site in the H gas ensemble at density $n_\text{H}$ becomes eligible for reaction. At T = 3000 K and T = 4000 K,  $t_N$ is 0.38 years and  1.78 years, respectively. 

Considering the larger time scale from $t_R$ and $t_N$, and for the H density ($n_\text{H}$, $n_{\text{H}^+}$ are $\sim 10^3$ cm$^{-3}$) of the early Universe at $z = 1100, 1500$, collision rates for the hard-sphere, $t_R$, with the energy constraint, KE $< 12$ cm$^{-1}$, would be $\sim 10^{-5}~\text{cm}^{-3} s^{-1}$. 

For comparison, the total recombination rate for three-body collisions between an H$^+$ ion and two H atoms is $\sim 10^{-22}$ cm$^{-3} s^{-1}$. Here, $K_3$ is $\approx 10^{-31}$ cm$^6 s^{-1}$ at T $\sim$ 1000 K \cite{cretu2022ion}. With non-adiabatic corrections, it could reach $\sim 10^{-18}$ cm$^{-3} s^{-1}$. It is, therefore, 17 (the usual) to 13 (with non-BO corrections for the JT-induced TTBR+CE reactions) orders of magnitude smaller than the rate for hard-sphere collisions at typical conditions.

For temperature T $\sim 1000$ K corresponding to $z \approx 1000$, and densities of H, H$^+$, and $e^-$ $\sim 10^3~\text{cm}^{-3}$, collision rates of other reactions of interest for H$_2$ formation in the early Universe, too, are many orders of magnitude smaller than the hard-sphere collision rates. 
For example, the first step of the H$_2^+$ mechanism (Eq. \ref{eq:eqH2a1}) has collision rate $\sim 10^{-11}$ cm$^{-3} s^{-1}$, (considering the rate coefficient of H$_2^+$ radiative association $\sim 10^{-17}$ cm$^3$s$^{-1}$ \cite{faure2024chemistry}). The first step of the H$^-$ mechanism (Eq. \ref{eq:eqH2b1}), it is $\sim 10^{-9}$ cm$^{-3} s^{-1}$, ( with rate coefficient of H$^-$ radiative attachment $\sim 10^{-15}$ cm$^3$s$^{-1}$ \cite{faure2024chemistry}). 

In addition, the time available for the JT-induced reactions leading to H$_2$ formation is very different from the thermalization time scales. Both $t_R$ and $t_N$ have durations on the order of years. In contrast, the interval between redshifts $z = 1100$ and $z = 1500$ is on the order of 10$^5$ years. This emphasizes that the thermalization processes under consideration occur on much shorter timescales than the overall cosmic evolution. 

This analysis demonstrates that the time required for the hydrogen gas ensemble to reach kinetic energies below $12$ cm$^{-1}$---thereby becoming effectively equivalent to an unconstrained ensemble for three-body reactions---is negligible compared to the timescales for H$_2$ formation in the early Universe. Consequently, the imposed energy constraint would not alter the probabilities of JT-induced reactions.\\

\noindent\textbf{Data availability}\\
The datasets generated during and/or analyzed during the study are available from the corresponding author upon reasonable request.\\

\noindent\textbf{Code availability}\\
MOLPRO software package for electronic structure calculations is publicly available \cite{MOLPRO-WIREs, Werner2020molpro}.

\putbib[references-H3+]
\end{bibunit}

\begin{thebibliography}{10}
\expandafter\ifx\csname url\endcsname\relax
  \def\url#1{\burl{#1}}\fi
\expandafter\ifx\csname urlprefix\endcsname\relax\def\urlprefix{URL }\fi
\providecommand{\bibinfo}[2]{#2}
\providecommand{\eprint}[2][]{\url{#2}}
\providecommand{\doi}[1]{\url{https://doi.org/#1}}
\bibcommenthead

\bibitem{barkana2001beginning}
\bibinfo{author}{Barkana, R.} \& \bibinfo{author}{Loeb, A.}
\newblock \bibinfo{title}{In the beginning: the first sources of light and the
  reionization of the {U}niverse}.
\newblock \emph{\bibinfo{journal}{Physics reports}}
  \textbf{\bibinfo{volume}{349}}, \bibinfo{pages}{125--238}
  (\bibinfo{year}{2001}).

\bibitem{bromm2009formation}
\bibinfo{author}{Bromm, V.}, \bibinfo{author}{Yoshida, N.},
  \bibinfo{author}{Hernquist, L.} \& \bibinfo{author}{McKee, C.~F.}
\newblock \bibinfo{title}{The formation of the first stars and galaxies}.
\newblock \emph{\bibinfo{journal}{Nature}} \textbf{\bibinfo{volume}{459}},
  \bibinfo{pages}{49--54} (\bibinfo{year}{2009}).

\bibitem{bromm2013formation}
\bibinfo{author}{Bromm, V.}
\newblock \bibinfo{title}{Formation of the first stars}.
\newblock \emph{\bibinfo{journal}{Reports on Progress in Physics}}
  \textbf{\bibinfo{volume}{76}}, \bibinfo{pages}{112901}
  (\bibinfo{year}{2013}).

\bibitem{tegmark1997small}
\bibinfo{author}{Tegmark, M.} \emph{et~al.}
\newblock \bibinfo{title}{How small were the first cosmological objects?}
\newblock \emph{\bibinfo{journal}{The Astrophysical Journal}}
  \textbf{\bibinfo{volume}{474}}, \bibinfo{pages}{1} (\bibinfo{year}{1997}).

\bibitem{klessen2023first}
\bibinfo{author}{Klessen, R.~S.} \& \bibinfo{author}{Glover, S.~C.}
\newblock \bibinfo{title}{The first stars: formation, properties, and impact}.
\newblock \emph{\bibinfo{journal}{Annual Review of Astronomy and Astrophysics}}
  \textbf{\bibinfo{volume}{61}}, \bibinfo{pages}{65--130}
  (\bibinfo{year}{2023}).

\bibitem{greif2015numerical}
\bibinfo{author}{Greif, T.~H.}
\newblock \bibinfo{title}{The numerical frontier of the high-redshift
  {U}niverse}.
\newblock \emph{\bibinfo{journal}{Computational Astrophysics and Cosmology}}
  \textbf{\bibinfo{volume}{2}}, \bibinfo{pages}{3} (\bibinfo{year}{2015}).

\bibitem{saslaw1967molecular}
\bibinfo{author}{Saslaw, W.~C.} \& \bibinfo{author}{Zipoy, D.}
\newblock \bibinfo{title}{Molecular hydrogen in pre-galactic gas clouds}.
\newblock \emph{\bibinfo{journal}{Nature}} \textbf{\bibinfo{volume}{216}},
  \bibinfo{pages}{976--978} (\bibinfo{year}{1967}).

\bibitem{peebles1968origin}
\bibinfo{author}{Peebles, P.} \& \bibinfo{author}{Dicke, R.}
\newblock \bibinfo{title}{Origin of the globular star clusters}.
\newblock \emph{\bibinfo{journal}{Astrophysical Journal, vol. 154, p. 891}}
  \textbf{\bibinfo{volume}{154}}, \bibinfo{pages}{891} (\bibinfo{year}{1968}).

\bibitem{galli2013dawn}
\bibinfo{author}{Galli, D.} \& \bibinfo{author}{Palla, F.}
\newblock \bibinfo{title}{The dawn of chemistry}.
\newblock \emph{\bibinfo{journal}{Annual Review of Astronomy and Astrophysics}}
  \textbf{\bibinfo{volume}{51}}, \bibinfo{pages}{163--206}
  (\bibinfo{year}{2013}).

\bibitem{bunker2023jades}
\bibinfo{author}{Bunker, A.~J.} \emph{et~al.}
\newblock \bibinfo{title}{{JADES} {NIRS}pec {S}pectroscopy of {GN}-z11:
  {L}yman-$\alpha$ emission and possible enhanced nitrogen abundance in a z =
  10.60 luminous galaxy}.
\newblock \emph{\bibinfo{journal}{Astronomy \& Astrophysics}}
  \textbf{\bibinfo{volume}{677}}, \bibinfo{pages}{A88} (\bibinfo{year}{2023}).

\bibitem{robertson2024earliest}
\bibinfo{author}{Robertson, B.} \emph{et~al.}
\newblock \bibinfo{title}{Earliest galaxies in the {JADES} origins field:
  {L}uminosity function and cosmic star formation rate density 300 {M}yr after
  the {B}ig {B}ang}.
\newblock \emph{\bibinfo{journal}{The Astrophysical Journal}}
  \textbf{\bibinfo{volume}{970}}, \bibinfo{pages}{31} (\bibinfo{year}{2024}).

\bibitem{carniani2024spectroscopic}
\bibinfo{author}{Carniani, S.} \emph{et~al.}
\newblock \bibinfo{title}{Spectroscopic confirmation of two luminous galaxies
  at a redshift of 14}.
\newblock \emph{\bibinfo{journal}{Nature}} \textbf{\bibinfo{volume}{633}},
  \bibinfo{pages}{318--322} (\bibinfo{year}{2024}).

\bibitem{zavala2025luminous}
\bibinfo{author}{Zavala, J.~A.} \emph{et~al.}
\newblock \bibinfo{title}{A luminous and young galaxy at z = 12.33 revealed by
  a {JWST/MIRI} detection of {H}$\alpha$ and [{O} iii]}.
\newblock \emph{\bibinfo{journal}{Nature Astronomy}}
  \textbf{\bibinfo{volume}{9}}, \bibinfo{pages}{155--164}
  (\bibinfo{year}{2025}).

\bibitem{naidu2025cosmic}
\bibinfo{author}{Naidu, R.~P.} \emph{et~al.}
\newblock \bibinfo{title}{A cosmic miracle: A remarkably luminous galaxy at
  $z_{\text{spec}}$ = 14.44 confirmed with {JWST}}.
\newblock \emph{\bibinfo{journal}{arXiv preprint arXiv:2505.11263}}
  (\bibinfo{year}{2025}).

\bibitem{adamo2025first}
\bibinfo{author}{Adamo, A.} \emph{et~al.}
\newblock \bibinfo{title}{The first billion years according to {JWST}}.
\newblock \emph{\bibinfo{journal}{Nature Astronomy}}
  \textbf{\bibinfo{volume}{9}}, \bibinfo{pages}{1134--1147}
  (\bibinfo{year}{2025}).

\bibitem{carniani2025eventful}
\bibinfo{author}{Carniani, S.} \emph{et~al.}
\newblock \bibinfo{title}{The eventful life of a luminous galaxy at z = 14:
  metal enrichment, feedback, and low gas fraction?}
\newblock \emph{\bibinfo{journal}{Astronomy \& Astrophysics}}
  \textbf{\bibinfo{volume}{696}}, \bibinfo{pages}{A87} (\bibinfo{year}{2025}).

\bibitem{xiao2024accelerated}
\bibinfo{author}{Xiao, M.} \emph{et~al.}
\newblock \bibinfo{title}{Accelerated formation of ultra-massive galaxies in
  the first billion years}.
\newblock \emph{\bibinfo{journal}{Nature}} \textbf{\bibinfo{volume}{635}},
  \bibinfo{pages}{311--315} (\bibinfo{year}{2024}).

\bibitem{perez2025rise}
\bibinfo{author}{P{\'e}rez-Gonz{\'a}lez, P.~G.} \emph{et~al.}
\newblock \bibinfo{title}{The rise of the galactic empire: Ultraviolet
  luminosity functions at z $\sim$ 17 and z $\sim$ 25 estimated with the
  {MIDIS}+{NGDEEP} {U}ltra-deep {JWST/NIRC}am data set}.
\newblock \emph{\bibinfo{journal}{The Astrophysical Journal}}
  \textbf{\bibinfo{volume}{991}}, \bibinfo{pages}{179} (\bibinfo{year}{2025}).

\bibitem{mason2023brightest}
\bibinfo{author}{Mason, C.~A.}, \bibinfo{author}{Trenti, M.} \&
  \bibinfo{author}{Treu, T.}
\newblock \bibinfo{title}{The brightest galaxies at cosmic dawn}.
\newblock \emph{\bibinfo{journal}{Monthly Notices of the Royal Astronomical
  Society}} \textbf{\bibinfo{volume}{521}}, \bibinfo{pages}{497--503}
  (\bibinfo{year}{2023}).

\bibitem{ferrara2023stunning}
\bibinfo{author}{Ferrara, A.}, \bibinfo{author}{Pallottini, A.} \&
  \bibinfo{author}{Dayal, P.}
\newblock \bibinfo{title}{On the stunning abundance of super-early, luminous
  galaxies revealed by {JWST}}.
\newblock \emph{\bibinfo{journal}{Monthly Notices of the Royal Astronomical
  Society}} \textbf{\bibinfo{volume}{522}}, \bibinfo{pages}{3986--3991}
  (\bibinfo{year}{2023}).

\bibitem{trinca2024exploring}
\bibinfo{author}{Trinca, A.} \emph{et~al.}
\newblock \bibinfo{title}{Exploring the nature of {UV}-bright z $\gtrsim$ 10
  galaxies detected by {JWST}: star formation, black hole accretion, or a
  non-universal {IMF}?}
\newblock \emph{\bibinfo{journal}{Monthly Notices of the Royal Astronomical
  Society}} \textbf{\bibinfo{volume}{529}}, \bibinfo{pages}{3563--3581}
  (\bibinfo{year}{2024}).

\bibitem{costantin2023milky}
\bibinfo{author}{Costantin, L.} \emph{et~al.}
\newblock \bibinfo{title}{A {M}ilky {W}ay-like barred spiral galaxy at a
  redshift of 3}.
\newblock \emph{\bibinfo{journal}{Nature}} \textbf{\bibinfo{volume}{623}},
  \bibinfo{pages}{499--501} (\bibinfo{year}{2023}).

\bibitem{wang2025giant}
\bibinfo{author}{Wang, W.} \emph{et~al.}
\newblock \bibinfo{title}{A giant disk galaxy two billion years after the {B}ig
  {B}ang}.
\newblock \emph{\bibinfo{journal}{Nature Astronomy}}
  \textbf{\bibinfo{volume}{9}}, \bibinfo{pages}{710--719}
  (\bibinfo{year}{2025}).

\bibitem{jain2025grand}
\bibinfo{author}{Jain, R.} \& \bibinfo{author}{Wadadekar, Y.}
\newblock \bibinfo{title}{A grand-design spiral galaxy 1.5 billion years after
  the {B}ig {B}ang with {JWST}}.
\newblock \emph{\bibinfo{journal}{Astronomy \& Astrophysics}}
  \textbf{\bibinfo{volume}{703}}, \bibinfo{pages}{A96} (\bibinfo{year}{2025}).

\bibitem{fan2023quasars}
\bibinfo{author}{Fan, X.}, \bibinfo{author}{Ba{\~n}ados, E.} \&
  \bibinfo{author}{Simcoe, R.~A.}
\newblock \bibinfo{title}{Quasars and the intergalactic medium at cosmic dawn}.
\newblock \emph{\bibinfo{journal}{Annual Review of Astronomy and Astrophysics}}
  \textbf{\bibinfo{volume}{61}}, \bibinfo{pages}{373--426}
  (\bibinfo{year}{2023}).

\bibitem{inayoshi2020assembly}
\bibinfo{author}{Inayoshi, K.}, \bibinfo{author}{Visbal, E.} \&
  \bibinfo{author}{Haiman, Z.}
\newblock \bibinfo{title}{The assembly of the first massive black holes}.
\newblock \emph{\bibinfo{journal}{Annual Review of Astronomy and Astrophysics}}
  \textbf{\bibinfo{volume}{58}}, \bibinfo{pages}{27--97}
  (\bibinfo{year}{2020}).

\bibitem{volonteri2021origins}
\bibinfo{author}{Volonteri, M.}, \bibinfo{author}{Habouzit, M.} \&
  \bibinfo{author}{Colpi, M.}
\newblock \bibinfo{title}{The origins of massive black holes}.
\newblock \emph{\bibinfo{journal}{Nature Reviews Physics}}
  \textbf{\bibinfo{volume}{3}}, \bibinfo{pages}{732--743}
  (\bibinfo{year}{2021}).

\bibitem{smith2019supermassive}
\bibinfo{author}{Smith, A.} \& \bibinfo{author}{Bromm, V.}
\newblock \bibinfo{title}{Supermassive black holes in the early {U}niverse}.
\newblock \emph{\bibinfo{journal}{Contemporary Physics}}
  (\bibinfo{year}{2019}).

\bibitem{jeon2025physical}
\bibinfo{author}{Jeon, J.}, \bibinfo{author}{Bromm, V.}, \bibinfo{author}{Liu,
  B.} \& \bibinfo{author}{Finkelstein, S.~L.}
\newblock \bibinfo{title}{Physical pathways for {JWST}-observed supermassive
  black holes in the early {U}niverse}.
\newblock \emph{\bibinfo{journal}{The Astrophysical Journal}}
  \textbf{\bibinfo{volume}{979}}, \bibinfo{pages}{127} (\bibinfo{year}{2025}).

\bibitem{hirata2006cosmological}
\bibinfo{author}{Hirata, C.~M.} \& \bibinfo{author}{Padmanabhan, N.}
\newblock \bibinfo{title}{Cosmological production of {H}$_2$ before the
  formation of the first galaxies}.
\newblock \emph{\bibinfo{journal}{Monthly Notices of the Royal Astronomical
  Society}} \textbf{\bibinfo{volume}{372}}, \bibinfo{pages}{1175--1186}
  (\bibinfo{year}{2006}).

\bibitem{grieco2023enhanced}
\bibinfo{author}{Grieco, F.}, \bibinfo{author}{Theul{\'e}, P.},
  \bibinfo{author}{De~Looze, I.} \& \bibinfo{author}{Dulieu, F.}
\newblock \bibinfo{title}{Enhanced star formation through the high-temperature
  formation of {H}$_2$ on carbonaceous dust grains}.
\newblock \emph{\bibinfo{journal}{Nature Astronomy}}
  \textbf{\bibinfo{volume}{7}}, \bibinfo{pages}{541--545}
  (\bibinfo{year}{2023}).

\bibitem{pandey2024ultracold}
\bibinfo{author}{Pandey, A.}, \bibinfo{author}{Vexiau, R.},
  \bibinfo{author}{Marcassa, L.~G.}, \bibinfo{author}{Dulieu, O.} \&
  \bibinfo{author}{Bouloufa-Maafa, N.}
\newblock \bibinfo{title}{Ultracold charged atom-dimer collisions:
  State-selective charge exchange and three-body recombination}.
\newblock \emph{\bibinfo{journal}{Physical Review Research}}
  \textbf{\bibinfo{volume}{6}}, \bibinfo{pages}{043010} (\bibinfo{year}{2024}).

\bibitem{yarkony1996diabolical}
\bibinfo{author}{Yarkony, D.~R.}
\newblock \bibinfo{title}{Diabolical conical intersections}.
\newblock \emph{\bibinfo{journal}{Reviews of Modern Physics}}
  \textbf{\bibinfo{volume}{68}}, \bibinfo{pages}{985} (\bibinfo{year}{1996}).

\bibitem{coppola2013non}
\bibinfo{author}{Coppola, C.~M.}, \bibinfo{author}{Galli, D.},
  \bibinfo{author}{Palla, F.}, \bibinfo{author}{Longo, S.} \&
  \bibinfo{author}{Chluba, J.}
\newblock \bibinfo{title}{Non-thermal photons and {H}$_2$ formation in the
  early {U}niverse}.
\newblock \emph{\bibinfo{journal}{Monthly Notices of the Royal Astronomical
  Society}} \textbf{\bibinfo{volume}{434}}, \bibinfo{pages}{114--122}
  (\bibinfo{year}{2013}).

\bibitem{ferretti1997quantum}
\bibinfo{author}{Ferretti, A.}, \bibinfo{author}{Lami, A.} \&
  \bibinfo{author}{Villani, G.}
\newblock \bibinfo{title}{Quantum dynamics of a model system with a conical
  intersection}.
\newblock \emph{\bibinfo{journal}{J. Chem. Phys.}}
  \textbf{\bibinfo{volume}{106}}, \bibinfo{pages}{934--941}
  (\bibinfo{year}{1997}).

\bibitem{farfan2020systematic}
\bibinfo{author}{Farfan, C.~A.} \& \bibinfo{author}{Turner, D.~B.}
\newblock \bibinfo{title}{A systematic model study quantifying how conical
  intersection topography modulates photochemical reactions}.
\newblock \emph{\bibinfo{journal}{Phys. Chem. Chem. Phys.}}
  \textbf{\bibinfo{volume}{22}}, \bibinfo{pages}{20265--20283}
  (\bibinfo{year}{2020}).

\bibitem{juodvzbalis2024rosetta}
\bibinfo{author}{Juod{\v{z}}balis, I.} \emph{et~al.}
\newblock \bibinfo{title}{{JADES}--the {R}osetta stone of {JWST}-discovered
  {AGN}: deciphering the intriguing nature of early {AGN}}.
\newblock \emph{\bibinfo{journal}{Monthly Notices of the Royal Astronomical
  Society}} \textbf{\bibinfo{volume}{535}}, \bibinfo{pages}{853--873}
  (\bibinfo{year}{2024}).

\bibitem{maiolino2024diverse}
\bibinfo{author}{Maiolino, R.} \emph{et~al.}
\newblock \bibinfo{title}{{JADES}-{T}he diverse population of infant black
  holes at 4 $<$ z $<$ 11: Merging, tiny, poor, but mighty}.
\newblock \emph{\bibinfo{journal}{Astronomy \& Astrophysics}}
  \textbf{\bibinfo{volume}{691}}, \bibinfo{pages}{A145} (\bibinfo{year}{2024}).

\bibitem{matthee2024little}
\bibinfo{author}{Matthee, J.} \emph{et~al.}
\newblock \bibinfo{title}{Little red dots: an abundant population of faint
  active galactic nuclei at z $\sim$ 5 revealed by the {EIGER} and {FRESCO}
  {JWST} surveys}.
\newblock \emph{\bibinfo{journal}{The Astrophysical Journal}}
  \textbf{\bibinfo{volume}{963}}, \bibinfo{pages}{129} (\bibinfo{year}{2024}).

\bibitem{taylor2025capers}
\bibinfo{author}{Taylor, A.~J.} \emph{et~al.}
\newblock \bibinfo{title}{{CAPERS-LRD}-z9: A gas-enshrouded {L}ittle {R}ed
  {D}ot hosting a broad-line {A}ctive {G}alactic {N}ucleus at z = 9.288}.
\newblock \emph{\bibinfo{journal}{The Astrophysical Journal Letters}}
  \textbf{\bibinfo{volume}{989}}, \bibinfo{pages}{L7} (\bibinfo{year}{2025}).

\bibitem{harrison2024observational}
\bibinfo{author}{Harrison, C.~M.} \& \bibinfo{author}{Ramos~Almeida, C.}
\newblock \bibinfo{title}{Observational tests of active galactic nuclei
  feedback: An overview of approaches and interpretation}.
\newblock \emph{\bibinfo{journal}{Galaxies}} \textbf{\bibinfo{volume}{12}},
  \bibinfo{pages}{17} (\bibinfo{year}{2024}).

\bibitem{alexander2025drives}
\bibinfo{author}{Alexander, D.} \emph{et~al.}
\newblock \bibinfo{title}{What drives the growth of black holes: A decade of
  progress}.
\newblock \emph{\bibinfo{journal}{New Astronomy Reviews}}
  \bibinfo{pages}{101733} (\bibinfo{year}{2025}).

\bibitem{riffel2020active}
\bibinfo{author}{Riffel, R.~A.}, \bibinfo{author}{Zakamska, N.~L.} \&
  \bibinfo{author}{Riffel, R.}
\newblock \bibinfo{title}{Active galactic nuclei winds as the origin of the
  {H}$_2$ emission excess in nearby galaxies}.
\newblock \emph{\bibinfo{journal}{Monthly Notices of the Royal Astronomical
  Society}} \textbf{\bibinfo{volume}{491}}, \bibinfo{pages}{1518--1529}
  (\bibinfo{year}{2020}).

\bibitem{gallagher2019widespread}
\bibinfo{author}{Gallagher, R.} \emph{et~al.}
\newblock \bibinfo{title}{Widespread star formation inside galactic outflows}.
\newblock \emph{\bibinfo{journal}{Monthly Notices of the Royal Astronomical
  Society}} \textbf{\bibinfo{volume}{485}}, \bibinfo{pages}{3409--3429}
  (\bibinfo{year}{2019}).

\bibitem{richings2018origin}
\bibinfo{author}{Richings, A.~J.} \& \bibinfo{author}{Faucher-Giguere, C.-A.}
\newblock \bibinfo{title}{The origin of fast molecular outflows in quasars:
  molecule formation in {AGN}-driven galactic winds}.
\newblock \emph{\bibinfo{journal}{Monthly Notices of the Royal Astronomical
  Society}} \textbf{\bibinfo{volume}{474}}, \bibinfo{pages}{3673--3699}
  (\bibinfo{year}{2018}).

\bibitem{ripamonti2007role}
\bibinfo{author}{Ripamonti, E.}
\newblock \bibinfo{title}{The role of {HD} cooling in primordial star
  formation}.
\newblock \emph{\bibinfo{journal}{Monthly Notices of the Royal Astronomical
  Society}} \textbf{\bibinfo{volume}{376}}, \bibinfo{pages}{709--718}
  (\bibinfo{year}{2007}).

\bibitem{mebane2018persistence}
\bibinfo{author}{Mebane, R.~H.}, \bibinfo{author}{Mirocha, J.} \&
  \bibinfo{author}{Furlanetto, S.~R.}
\newblock \bibinfo{title}{The persistence of {P}opulation {III} star
  formation}.
\newblock \emph{\bibinfo{journal}{Monthly Notices of the Royal Astronomical
  Society}} \textbf{\bibinfo{volume}{479}}, \bibinfo{pages}{4544--4559}
  (\bibinfo{year}{2018}).

\bibitem{madau2001massive}
\bibinfo{author}{Madau, P.} \& \bibinfo{author}{Rees, M.~J.}
\newblock \bibinfo{title}{Massive black holes as population {III} remnants}.
\newblock \emph{\bibinfo{journal}{The Astrophysical Journal}}
  \textbf{\bibinfo{volume}{551}}, \bibinfo{pages}{L27} (\bibinfo{year}{2001}).

\bibitem{lupi2024sustained}
\bibinfo{author}{Lupi, A.}, \bibinfo{author}{Quadri, G.},
  \bibinfo{author}{Volonteri, M.}, \bibinfo{author}{Colpi, M.} \&
  \bibinfo{author}{Regan, J.~A.}
\newblock \bibinfo{title}{Sustained super-{E}ddington accretion in
  high-redshift quasars}.
\newblock \emph{\bibinfo{journal}{Astronomy \& Astrophysics}}
  \textbf{\bibinfo{volume}{686}}, \bibinfo{pages}{A256} (\bibinfo{year}{2024}).

\bibitem{johnson2007aftermath}
\bibinfo{author}{Johnson, J.~L.} \& \bibinfo{author}{Bromm, V.}
\newblock \bibinfo{title}{The aftermath of the first stars: massive black
  holes}.
\newblock \emph{\bibinfo{journal}{Monthly Notices of the Royal Astronomical
  Society}} \textbf{\bibinfo{volume}{374}}, \bibinfo{pages}{1557--1568}
  (\bibinfo{year}{2007}).

\bibitem{o2025predicting}
\bibinfo{author}{O'Brennan, H.} \emph{et~al.}
\newblock \bibinfo{title}{Predicting the number density of heavy seed massive
  black holes due to an intense {L}yman-{W}erner field}.
\newblock \emph{\bibinfo{journal}{arXiv preprint arXiv:2502.00574}}
  (\bibinfo{year}{2025}).

\bibitem{suh2025super}
\bibinfo{author}{Suh, H.} \emph{et~al.}
\newblock \bibinfo{title}{A super-{E}ddington-accreting black hole $\sim$ 1.5
  {G}yr after the {B}ig {B}ang observed with {JWST}}.
\newblock \emph{\bibinfo{journal}{Nature Astronomy}}
  \textbf{\bibinfo{volume}{9}}, \bibinfo{pages}{271--279}
  (\bibinfo{year}{2025}).

\bibitem{ighina2025x}
\bibinfo{author}{Ighina, L.} \emph{et~al.}
\newblock \bibinfo{title}{X-{R}ay investigation of possible super-{E}ddington
  accretion in a radio-loud quasar at z = 6.13}.
\newblock \emph{\bibinfo{journal}{The Astrophysical Journal Letters}}
  \textbf{\bibinfo{volume}{990}}, \bibinfo{pages}{L56} (\bibinfo{year}{2025}).

\bibitem{geris2026jades}
\bibinfo{author}{Geris, S.} \emph{et~al.}
\newblock \bibinfo{title}{{JADES} reveals a large population of low-mass black
  holes at high redshift}.
\newblock \emph{\bibinfo{journal}{Monthly Notices of the Royal Astronomical
  Society}} \textbf{\bibinfo{volume}{545}}, \bibinfo{pages}{staf1979}
  (\bibinfo{year}{2026}).

\bibitem{maiolino2024small}
\bibinfo{author}{Maiolino, R.} \emph{et~al.}
\newblock \bibinfo{title}{A small and vigorous black hole in the early
  {U}niverse}.
\newblock \emph{\bibinfo{journal}{Nature}} \textbf{\bibinfo{volume}{627}},
  \bibinfo{pages}{59--63} (\bibinfo{year}{2024}).

\bibitem{reines2015relations}
\bibinfo{author}{Reines, A.~E.} \& \bibinfo{author}{Volonteri, M.}
\newblock \bibinfo{title}{Relations between central black hole mass and total
  galaxy stellar mass in the local {U}niverse}.
\newblock \emph{\bibinfo{journal}{The Astrophysical Journal}}
  \textbf{\bibinfo{volume}{813}}, \bibinfo{pages}{82} (\bibinfo{year}{2015}).

\end{thebibliography}


\begin{thebibliography}{10}
\expandafter\ifx\csname url\endcsname\relax
  \def\url#1{\burl{#1}}\fi
\expandafter\ifx\csname urlprefix\endcsname\relax\def\urlprefix{URL }\fi
\providecommand{\bibinfo}[2]{#2}
\providecommand{\eprint}[2][]{\url{#2}}
\providecommand{\doi}[1]{\url{https://doi.org/#1}}
\bibcommenthead

\bibitem{MOLPRO-WIREs}
\bibinfo{author}{Werner, H.-J.}, \bibinfo{author}{Knowles, P.~J.},
  \bibinfo{author}{Knizia, G.}, \bibinfo{author}{Manby, F.~R.} \&
  \bibinfo{author}{Sch{\"u}tz, M.}
\newblock \bibinfo{title}{{Molpro: a general{-}purpose quantum chemistry
  program package}}.
\newblock \emph{\bibinfo{journal}{WIREs Comput Mol Sci}}
  \textbf{\bibinfo{volume}{2}}, \bibinfo{pages}{242--253}
  (\bibinfo{year}{2012}).

\bibitem{werner1988efficient}
\bibinfo{author}{Werner, H.-J.} \& \bibinfo{author}{Knowles, P.~J.}
\newblock \bibinfo{title}{An efficient internally contracted
  multiconfiguration--reference configuration interaction method}.
\newblock \emph{\bibinfo{journal}{The Journal of chemical physics}}
  \textbf{\bibinfo{volume}{89}}, \bibinfo{pages}{5803--5814}
  (\bibinfo{year}{1988}).

\bibitem{viegas2007accurate}
\bibinfo{author}{Viegas, L.~P.}, \bibinfo{author}{Alijah, A.} \&
  \bibinfo{author}{Varandas, A.~J.}
\newblock \bibinfo{title}{Accurate ab initio based multisheeted double
  many-body expansion potential energy surface for the three lowest electronic
  singlet states of {H}$_3^+$}.
\newblock \emph{\bibinfo{journal}{The Journal of chemical physics}}
  \textbf{\bibinfo{volume}{126}} (\bibinfo{year}{2007}).

\bibitem{aguado2021three}
\bibinfo{author}{Aguado, A.}, \bibinfo{author}{Roncero, O.} \&
  \bibinfo{author}{Sanz-Sanz, C.}
\newblock \bibinfo{title}{Three states global fittings with improved long
  range: singlet and triplet states of {H}$_3^+$}.
\newblock \emph{\bibinfo{journal}{Physical Chemistry Chemical Physics}}
  \textbf{\bibinfo{volume}{23}}, \bibinfo{pages}{7735--7747}
  (\bibinfo{year}{2021}).

\bibitem{pandey2024ultracold}
\bibinfo{author}{Pandey, A.}, \bibinfo{author}{Vexiau, R.},
  \bibinfo{author}{Marcassa, L.~G.}, \bibinfo{author}{Dulieu, O.} \&
  \bibinfo{author}{Bouloufa-Maafa, N.}
\newblock \bibinfo{title}{Ultracold charged atom-dimer collisions:
  State-selective charge exchange and three-body recombination}.
\newblock \emph{\bibinfo{journal}{Physical Review Research}}
  \textbf{\bibinfo{volume}{6}}, \bibinfo{pages}{043010} (\bibinfo{year}{2024}).

\bibitem{cretu2022ion}
\bibinfo{author}{Cretu, M.~T.}, \bibinfo{author}{Mirahmadi, M.} \&
  \bibinfo{author}{P{\'e}rez-R{\'\i}os, J.}
\newblock \bibinfo{title}{Ion-atom-atom three-body recombination in cold
  hydrogen and deuterium plasmas}.
\newblock \emph{\bibinfo{journal}{Physical Review A}}
  \textbf{\bibinfo{volume}{106}}, \bibinfo{pages}{023316}
  (\bibinfo{year}{2022}).

\bibitem{mirahmadi2023ion}
\bibinfo{author}{Mirahmadi, M.} \& \bibinfo{author}{P{\'e}rez-R{\'\i}os, J.}
\newblock \bibinfo{title}{Ion-atom-atom three-body recombination: From the cold
  to the thermal regime}.
\newblock \emph{\bibinfo{journal}{The Journal of Chemical Physics}}
  \textbf{\bibinfo{volume}{158}} (\bibinfo{year}{2023}).

\bibitem{saslaw1967molecular}
\bibinfo{author}{Saslaw, W.~C.} \& \bibinfo{author}{Zipoy, D.}
\newblock \bibinfo{title}{Molecular hydrogen in pre-galactic gas clouds}.
\newblock \emph{\bibinfo{journal}{Nature}} \textbf{\bibinfo{volume}{216}},
  \bibinfo{pages}{976--978} (\bibinfo{year}{1967}).

\bibitem{peebles1968origin}
\bibinfo{author}{Peebles, P.} \& \bibinfo{author}{Dicke, R.}
\newblock \bibinfo{title}{Origin of the globular star clusters}.
\newblock \emph{\bibinfo{journal}{Astrophysical Journal, vol. 154, p. 891}}
  \textbf{\bibinfo{volume}{154}}, \bibinfo{pages}{891} (\bibinfo{year}{1968}).

\bibitem{mcdowell1961formation}
\bibinfo{author}{McDowell, M.}
\newblock \bibinfo{title}{On the formation of {H}$_2$ in {HI} regions}.
\newblock \emph{\bibinfo{journal}{The Observatory, Vol. 81, p. 240-243 (1961)}}
  \textbf{\bibinfo{volume}{81}}, \bibinfo{pages}{240--243}
  (\bibinfo{year}{1961}).

\bibitem{bieniek1979associative}
\bibinfo{author}{Bieniek, R.~J.} \& \bibinfo{author}{Dalgarno, A.}
\newblock \bibinfo{title}{Associative detachment in collisions of {H} and
  {H}$^-$}.
\newblock \emph{\bibinfo{journal}{Astrophysical Journal, Part 1, vol. 228, Mar.
  1, 1979, p. 635-639.}} \textbf{\bibinfo{volume}{228}},
  \bibinfo{pages}{635--639} (\bibinfo{year}{1979}).

\bibitem{hirata2006cosmological}
\bibinfo{author}{Hirata, C.~M.} \& \bibinfo{author}{Padmanabhan, N.}
\newblock \bibinfo{title}{Cosmological production of {H}$_2$ before the
  formation of the first galaxies}.
\newblock \emph{\bibinfo{journal}{Monthly Notices of the Royal Astronomical
  Society}} \textbf{\bibinfo{volume}{372}}, \bibinfo{pages}{1175--1186}
  (\bibinfo{year}{2006}).

\bibitem{sugimura2016role}
\bibinfo{author}{Sugimura, K.}, \bibinfo{author}{Coppola, C.~M.},
  \bibinfo{author}{Omukai, K.}, \bibinfo{author}{Galli, D.} \&
  \bibinfo{author}{Palla, F.}
\newblock \bibinfo{title}{Role of the {H}$_2^+$ channel in the primordial star
  formation under strong radiation field and the critical intensity for the
  supermassive star formation}.
\newblock \emph{\bibinfo{journal}{Monthly Notices of the Royal Astronomical
  Society}} \textbf{\bibinfo{volume}{456}}, \bibinfo{pages}{270--277}
  (\bibinfo{year}{2016}).

\bibitem{ghosh2017beyond}
\bibinfo{author}{Ghosh, S.} \emph{et~al.}
\newblock \bibinfo{title}{Beyond {B}orn-{O}ppenheimer theory for ab initio
  constructed diabatic potential energy surfaces of singlet {H}$_3^+$ to study
  reaction dynamics using coupled {3D} time-dependent wave-packet approach}.
\newblock \emph{\bibinfo{journal}{The Journal of Chemical Physics}}
  \textbf{\bibinfo{volume}{147}}, \bibinfo{pages}{074105}
  (\bibinfo{year}{2017}).

\bibitem{mukherjee2019beyond}
\bibinfo{author}{Mukherjee, B.} \emph{et~al.}
\newblock \bibinfo{title}{Beyond {B}orn-{O}ppenheimer theory for spectroscopic
  and scattering processes}.
\newblock \emph{\bibinfo{journal}{International Reviews in Physical Chemistry}}
  \textbf{\bibinfo{volume}{38}}, \bibinfo{pages}{287--341}
  (\bibinfo{year}{2019}).

\bibitem{ghosh2021charge}
\bibinfo{author}{Ghosh, S.}, \bibinfo{author}{Sahoo, T.},
  \bibinfo{author}{Baer, M.} \& \bibinfo{author}{Adhikari, S.}
\newblock \bibinfo{title}{Charge transfer processes for {H}$+${H}$_2^+$
  reaction employing coupled {3D} wavepacket approach on beyond
  {B}orn-{O}ppenheimer based ab initio constructed diabatic potential energy
  surfaces}.
\newblock \emph{\bibinfo{journal}{The Journal of Physical Chemistry A}}
  \textbf{\bibinfo{volume}{125}}, \bibinfo{pages}{731--745}
  (\bibinfo{year}{2021}).

\bibitem{sanz2021near}
\bibinfo{author}{Sanz-Sanz, C.}, \bibinfo{author}{Aguado, A.} \&
  \bibinfo{author}{Roncero, O.}
\newblock \bibinfo{title}{Near-resonant effects in the quantum dynamics of the
  {H} $+$ {H}$_2^+$ $\rightarrow$ {H}$_2$ $+$ {H}$^+$ charge transfer reaction
  and isotopic variants}.
\newblock \emph{\bibinfo{journal}{The Journal of Chemical Physics}}
  \textbf{\bibinfo{volume}{154}} (\bibinfo{year}{2021}).

\bibitem{roncero2022vibrational}
\bibinfo{author}{Roncero, O.}, \bibinfo{author}{Andrianarijaona, V.},
  \bibinfo{author}{Aguado, A.} \& \bibinfo{author}{Sanz-Sanz, C.}
\newblock \bibinfo{title}{Vibrational effects in the quantum dynamics of the
  {H} $+$ {D}$_2^+$ charge transfer reaction}.
\newblock \emph{\bibinfo{journal}{Molecular Physics}}
  \textbf{\bibinfo{volume}{120}}, \bibinfo{pages}{e1948125}
  (\bibinfo{year}{2022}).

\bibitem{krstic2002inelastic}
\bibinfo{author}{Krsti{\'c}, P.~S.}
\newblock \bibinfo{title}{Inelastic processes from vibrationally excited states
  in slow {H}$^+$ $+$ {H}$_2$ and {H} $+$ {H}$_2^+$ collisions: Excitations and
  charge transfer}.
\newblock \emph{\bibinfo{journal}{Physical Review A}}
  \textbf{\bibinfo{volume}{66}}, \bibinfo{pages}{042717}
  (\bibinfo{year}{2002}).

\bibitem{krstic2005vibrationally}
\bibinfo{author}{Krsti{\'c}, P.~S.}
\newblock \bibinfo{title}{Vibrationally resolved collisions in cold hydrogen
  plasma}.
\newblock \emph{\bibinfo{journal}{Nuclear Instruments and Methods in Physics
  Research Section B: Beam Interactions with Materials and Atoms}}
  \textbf{\bibinfo{volume}{241}}, \bibinfo{pages}{58--62}
  (\bibinfo{year}{2005}).

\bibitem{krstic2003three}
\bibinfo{author}{Krsti{\'c}, P.}, \bibinfo{author}{Janev, R.} \&
  \bibinfo{author}{Schultz, D.}
\newblock \bibinfo{title}{Three-body, diatomic association in cold hydrogen
  plasmas}.
\newblock \emph{\bibinfo{journal}{Journal of Physics B: Atomic, Molecular and
  Optical Physics}} \textbf{\bibinfo{volume}{36}}, \bibinfo{pages}{L249}
  (\bibinfo{year}{2003}).

\bibitem{faure2024chemistry}
\bibinfo{author}{Faure, A.}, \bibinfo{author}{Hily-Blant, P.},
  \bibinfo{author}{Pineau~des For{\^e}ts, G.} \& \bibinfo{author}{Flower, D.}
\newblock \bibinfo{title}{The chemistry and excitation of {H}$_2$ and {HD} in
  the early {U}niverse}.
\newblock \emph{\bibinfo{journal}{Monthly Notices of the Royal Astronomical
  Society}} \textbf{\bibinfo{volume}{531}}, \bibinfo{pages}{340--354}
  (\bibinfo{year}{2024}).

\bibitem{lykke1991threshold}
\bibinfo{author}{Lykke, K.}, \bibinfo{author}{Murray, K.} \&
  \bibinfo{author}{Lineberger, W.}
\newblock \bibinfo{title}{Threshold photodetachment of {H}$^-$}.
\newblock \emph{\bibinfo{journal}{Physical Review A}}
  \textbf{\bibinfo{volume}{43}}, \bibinfo{pages}{6104} (\bibinfo{year}{1991}).

\bibitem{zammit2017state}
\bibinfo{author}{Zammit, M.~C.} \emph{et~al.}
\newblock \bibinfo{title}{State-resolved photodissociation and radiative
  association data for the molecular hydrogen ion}.
\newblock \emph{\bibinfo{journal}{The Astrophysical Journal}}
  \textbf{\bibinfo{volume}{851}}, \bibinfo{pages}{64} (\bibinfo{year}{2017}).

\bibitem{galli1998chemistry}
\bibinfo{author}{Galli, D.} \& \bibinfo{author}{Palla, F.}
\newblock \bibinfo{title}{The chemistry of the early {U}niverse}.
\newblock \emph{\bibinfo{journal}{Astron. Astrophys}}
  \textbf{\bibinfo{volume}{335}}, \bibinfo{pages}{403--420}
  (\bibinfo{year}{1998}).

\bibitem{Werner2020molpro}
\bibinfo{author}{Werner, H.-J.} \emph{et~al.}
\newblock \bibinfo{title}{The molpro quantum chemistry package}.
\newblock \emph{\bibinfo{journal}{The Journal of Chemical Physics}}
  \textbf{\bibinfo{volume}{152}}, \bibinfo{pages}{144107}
  (\bibinfo{year}{2020}).
\newblock \urlprefix\url{https://doi.org/10.1063/5.0005081}.

\end{thebibliography}

\vspace{20pt}
\noindent\textbf{Acknowledgments} We acknowledge E. Roueff (LUX, Observatoire de Paris, France) and S. Charlot (Institut d'Astrophysique de Paris, France) for helpful discussions and critical comments; we greatly appreciate their input. 
We also acknowledge support from the Agence Nationale de la Recherche, Grant No. ANR-21-CE30-0060-01 (COCOTRAMOS project) for the primary work on the rubidium system.\\

\noindent\textbf{Author Contributions} A.P., O.D., and N.B. conceived the study of molecular formation via Jahn-Teller-induced coupling. A.P. performed the calculations and led the drafting of the manuscript, with contributions to the text and figures from OD and NB. All authors contributed to the editing and revision of the manuscript.\\

\noindent\textbf{Competing interests} The authors declare no competing interests. \\

\noindent\textbf{Additional information}\\
Correspondence and requests for materials should be addressed to Amrendra Pandey.

\end{document}